\pdfoutput=1
\documentclass[12pt]{scrartcl}

\usepackage{preamble_W}
\usepackage{multirow}

\newcommand{\sys}[1]{\textsc{#1}}

\def\S{\ensuremath{\mathrm{S}}}
\def\I{\ensuremath{\mathrm{I}}}
\def\R{\ensuremath{\mathrm{R}}}
 \def\ga{\gamma}
 \def\Rn{{\cal R}_0}
 \def\ro{\Rn}
 \newcommand\be{\begin{equation}\label}
 \newcommand\ee{\end{equation}}
\newcommand{\defeq}{\coloneqq}

\newcommand{\differential}[1]{\mathrm{d} #1}

\newcommand{\eqstop}{.}
\newcommand{\eqcomma}{,}

\newcommand{\prob}{\mathsf{P}}
\newcommand{\probOf}[1]{\prob(#1)}

\newcommand{\myExp}[1]{\exp \bigl( #1 \bigr)  }

\newcommand{\ie}{\textit{i.e.}}

\newcommand{\myhighlight}[1]{\textcolor{tud9d}{\textbf{#1}}}

\renewcommand {\theequation}{\arabic{section}.\arabic{equation}}

\title{Survival Dynamical Systems for the Population-level Analysis of Epidemics}

\author{\color{tud1d}\sffamily Wasiur~R.~KhudaBukhsh\footnote{Mathematical Biosciences Institute, The Ohio State University, USA,
		email:\href{mailto:khudabukhsh.2@osu.edu}{khudabukhsh.2@osu.edu}  }      \hspace{1.5mm}\href{https://orcid.org/0000-0003-1803-0470}{\includegraphics[width=3mm]{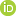}}, \emph{\small The Ohio State University, USA} \hfill 	 \\
    \color{tud1d}\sffamily Boseung~Choi\footnote{Division of Economics and Statistics, Department of National Statistics, Korea University Sejong campus,
		email:\href{mailto:cbskust@korea.ac.kr}{cbskust@korea.ac.kr}  }      \hspace{1.5mm}\href{https://orcid.org/0000-0001-7512-118X}{\includegraphics[width=3mm]{Figures/ORCID-iD_icon-16x16}}, \emph{\small Korea University Sejong Campus, Korea} \hfill 	 \\
    \color{tud1d}\sffamily Eben~Kenah\footnote{Division of Biostatistics, College of Public Health, The Ohio State University, USA,
		email:\href{mailto:kenah.1@osu.edu}{kenah.1@osu.edu}  }, \emph{\small The Ohio State University, USA} \hfill 	 \\
    \color{tud1d}\sffamily Grzegorz~A.~Rempa{\l}a\footnote{Division of Biostatistics, College of Public Health and Mathematical Biosciences Institute, The Ohio State University, USA,
		email:\href{mailto:rempala.3@osu.edu}{rempala.3@osu.edu}  }      \hspace{1.5mm}\href{https://orcid.org/0000-0002-6307-4555}{\includegraphics[width=3mm]{Figures/ORCID-iD_icon-16x16}}, \emph{\small The Ohio State University, USA} \hfill
    }
\date{}

\begin{document}
\maketitle


\begin{abstract} 
Motivated by the classical  \ac{SIR} epidemic models proposed by Kermack and Mckendrick, we consider a class of stochastic compartmental dynamical systems with a notion of partial ordering among the compartments. We call such systems unidirectional \acp{MTM}. We show that there is a natural way of interpreting a uni-directional \ac{MTM} as a \ac{SDS} that is described in terms of survival functions instead of population counts. This \ac{SDS} interpretation allows us to employ tools from survival analysis to address various issues with data collection and statistical inference of unidirectional \acp{MTM}. In particular, we propose and numerically validate a statistical inference procedure based on \ac{SDS}-likelihoods. We use the  \ac{SIR} model as a running example throughout the paper to illustrate the ideas. 
\end{abstract}



\section{Introduction}
\label{sec:intro}
\begin{figure}[ht]
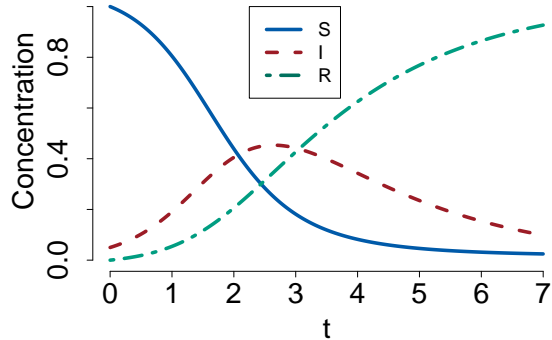

  \centering
  \includegraphics[scale=0.3]{{{SIR_curves_as_MTM}}}
  \caption{\label{fig:sir_ode_solution}%
  Survival analysis interpretation of  the  \ac{SIR} curves. We interpret the $S_t$ curve as the (right) tail distribution of the infection time of a susceptible individual: $S_t = \probOf{T_I > t}$ where $T_I$ is the transfer time of an individual from the susceptible to the infected compartment. The $R_t$ curve, upon multiplication with $\ro$, gives the growing cumulative hazard. Finally, the convolution of the infection time~$T_I$ and the infectious time~$T_R$ (time spent in the infected compartment) is given by the $I_t$ curve, after adjustment for the initially infecteds. Parameter values: $\beta=2, \gamma=0.5$ with initial condition $S_0=1, I_0=0.05, R_0=0$.
   }
\end{figure}

One of the earliest works in compartmental disease modeling is the seminal 1927 paper by Kermack and McKendrick~\cite{kermack1927contribution}. It  assumes the population is segregated into susceptible (S), infected (I), and recovered or removed (R) compartments.  Kermack and McKendrick proposed the following well known system of \acp{ODE} to describe the time evolution of the population proportions in each compartment, denoted by $S_t, I_t,$ and $R_t$ respectively:
 \begin{align}
 \begin{aligned}
 \dot{S}_t = {}&  -\beta S_t I_t \eqcomma    \\
 \dot{I}_t = {}& \beta S_t I_t - \gamma I_t \eqcomma    \\
 \dot{R}_t ={}& \gamma I_t  \eqstop
 \end{aligned} \label{eq:SIR_ODE}
 \end{align}
Here $\beta$ and $\gamma$ are the infection and recovery rates, respectively. 
Solutions to \Cref{eq:SIR_ODE} are often called the \ac{SIR} curves (see \Cref{fig:sir_ode_solution}).  In the absence of any specific contact structure, the law of mass action has been implicitly assumed, so an infectious individual can potentially infect any susceptible individual. Despite its popularity and widespread use over decades, the \ac{ODE} model in \Cref{eq:SIR_ODE} averages out individual dynamics and, therefore, does not capture the stochastic fluctuation of epidemic processes in real life. In particular, the practical problems of applying  \Cref{eq:SIR_ODE}  to data are:
 \begin{enumerate}
    \item \label{problem1} \textbf{Population size} Since the quantities in the \ac{SIR} equations are proportions, it is not immediately  clear how to apply them to real epidemics, which occur in  \emph{finite} susceptible populations.  Moreover, the size of the population is  often unknown.
   \item \label{problem4} \textbf{Likelihood} Since the \ac{SIR} equations are deterministic,  we cannot write a likelihood for epidemic data without further, often ad-hoc, statistical assumptions.
   \item \label{problem2} \textbf{Aggregation over individuals} The \ac{SIR} model represents the mean-field equations for (scaled) population counts, aggregating out individual characteristics.
     \item \label{problem3} \textbf{Aggregation over time} The real data are typically aggregated not just over the population but also over observed time periods, leading to  interval  censoring that cannot be easily incorporated into the \ac{SIR} equations.
 \end{enumerate}
The objective of this paper is to introduce a new way of interpreting \ac{SIR}  \Cref{eq:SIR_ODE} in terms of a survival function instead of population counts. This will address the first two problems directly, and it will also give us a theoretical foundation for addressing the remaining  two problems. Our approach will be applicable not only to mass-acton based \ac{SIR}-type  models but also  to a broad class of network-based epidemic models.

The \ac{SIR}
\Cref{eq:SIR_ODE}  is the  simplest  example of an epidemiological \ac{CoM}.
For the purpose of this paper,  we understand  a  \acs{CoM}  as a discrete set of states (compartments) paired with  a set of continuous  or discrete transition rules between them.   Continuous deterministic \ac{CoM} are often considered to describe the {\em macro} (population) level, as in \Cref{eq:SIR_ODE}.   Discrete stochastic  \acp{CoM}  are often considered to describe the  {\em micro} (individual) level dynamics of an epidemic.
 As we  outline below, there exists an interesting and so far largely unexplored  connection  between  the macro- and micro- level descriptions of \acp{CoM} (including the \ac{SIR} \Cref{eq:SIR_ODE}).   This connection is based on the notion of  mass transfer defined below, and it  gives us new insights into how to  address the practical problems of statistical inference listed above. The following definition will be useful:
\begin{myDefinition}[\acl{MTM}]
{Any compartmental model with mass conservation (\ie, constant total count or density of individuals) is termed a \acf{MTM}. In particular, when there exists a partial ordering on the set of  compartments, we call it a {unidirectional} \ac{MTM}}.  \QEDB
\end{myDefinition}


Following the above definition, it is clear that the \ac{SIR} model described by \Cref{eq:SIR_ODE} can be thought of as a unidirectional \ac{MTM}. In this paper, we shall show that this \ac{MTM} interpretation of the  \ac{SIR} \Cref{eq:SIR_ODE} can be used to describe the fate of an individual that starts in the first (susceptible) compartment and moves (with certain probability) to subsequent compartments\footnote{Recall that  there is an inherent partial ordering on the set of   compartments in a unidirectional \ac{MTM}.}. In other words, simple algebraic manipulation of the \ac{MTM} uncovers a precise description of  the survival dynamics of an individual  (see \Cref{fig:sir_ode_solution}). To emphasize this connection,  we shall often refer to a dynamical system describing unidirectional \ac{MTM} as a Survival Dynamical System (\acs{SDS}).

Our contributions in this paper can be summarized as follows: We propose a new way of describing unidirectional \acp{MTM} such as the \ac{SIR} \Cref{eq:SIR_ODE} in terms of a population survival function instead of population counts. This new interpretation will not only allow us to apply  all of the standard survival analysis tools to typical epidemic data, but  it will also address all four of the practical problems listed earlier. In particular,  based on the \ac{SDS} interpretation of unidirectional \acp{MTM}, we propose a simplified likelihood, called the \ac{SDS}-likelihood, for the purpose of statistical inference. We then  numerically verify  on simulation data examples that our new inference  method compares favorably with standard approaches  based on posterior likelihood and  \ac{MCMC} schemes.

The rest of the paper is structured as follows. \Cref{sec:background} briefly reviews the relevant  background on  mathematical modeling in epidemiological literature. In  \Cref{sec:mtm} we make  the \ac{MTM} interpretation of the \ac{SIR} \Cref{eq:SIR_ODE} precise whereas  in \Cref{sec:inference} we compare the standard as well as our proposed parameter estimation methods followed by numerical results in \Cref{sec:numerical}. Finally, we conclude the paper with a brief discussion in \Cref{sec:conclusion}. Additional mathematical preliminaries, statistical inference results and other supplementary material are provided in \Cref{appendix:prelims,appendix:inference,appendix:acronym}.



\setcounter{equation}{0}
\section{Background}
\label{sec:background}

%

Suppose we have $n$ susceptible and $m$ infectious individuals. Infectious individuals infect susceptible individuals, who change state from susceptible to infected. Infected individuals recover after an exponential infectious period. For the $i$-th individual, define the process $S_i$ such that $S_i(t) = 1$ if he or she is in the susceptible compartment at time $t$ and $S_i(t) = 0$ otherwise. Similarly, define the processes $I_i$ for the infected compartment and $R_i$ for the recovered compartment. Naturally, $S_i(t) + I_i(t) + R_i(t)  = 1$. We assume Markovianness throughout the course of the paper. That is, we assume $\{(S_i(t), I_i(t), R_i(t) )\}_{i=1,2,\ldots,n,n+1,\ldots,n+m; \; t\in [0,T]}$, for some $T \in (0,\infty)$, is a \ac{CTMC}. For the sake of notational convenience,  we have  labeled the initially susceptible individuals $1,2,\ldots,n$ and the initially infectious individuals $n+1,n+2,\ldots,n+m$.   Then, following the random time change representation of a \ac{CTMC} (see \cite[Chapter~6]{ethier2009markov}, \cite[Chapter~5]{britton2000stochEpi}), we can write, for $i=1,2,\ldots,n+m$,
\begin{align}
  \begin{aligned}
S_i(t) ={}& S_i(0) - Y_{i} \Big( \int_0^t \dfrac{\beta}{n}  S_i(s) \sum_{j=1}^{n+m} I_j(s) \, \differential{s}  \Big) \eqcomma \\
I_i(t) ={}& I_i(0) + Y_{i} \Big( \int_0^t \dfrac{\beta}{n}  S_i(s) \sum_{j=1}^{n+m} I_j(s) \, \differential{s}  \Big) - Z_{i} \Big( \int_0^t \gamma I_i(s)\, \differential{s}\Big) \eqcomma \\
R_i(t) ={} & Z_{i} \Big( \int_0^t \gamma I_i(s)\, \differential{s}\Big) \eqcomma
  \end{aligned}
  \label{eq:micro_RTCM_stoch_equations}
\end{align}
 where $Y_1,Y_2,\ldots,Y_{n+m}, Z_1, Z_2, \ldots, Z_{n+m}$ are independent unit-rate Poisson processes. 
 Models of this form are often called {\em agent-based models} in the literature \cite{khudabukhsh2018Lumpability,banisch2016agent} and if required, may be explicitly simulated by means of the so-called Doob-Gillespie algorithm \cite{gillespie1977exact,anderson2015stochastic,wilkinson2006stochastic}.

\subsection{Sellke construction}
\label{sec:background:Sellke}
An alternative construction of the micro model from a survival analysis perspective was  proposed by Sellke \cite{sellke1983asymptotic} as outlined  below.  Note that conditionally on the history  of the infection process $I$ (population count of infected) up to time $t$, the infection time $T_{i,I}$ of a susceptible individual $i$ is given by
\begin{align}
\probOf{T_{i,I} >t \mid  \left( I(s)  \right)_{s \in [0,t]}  } = {} & \myExp{ - \frac{\beta}{n} \int_0^t I(s) \, \differential{s}} \,  \eqstop \label{eq:infection_time_distribution}
\end{align}
Once a susceptible individual gets infected, he/she recovers after an infectious period that follows an exponential distribution with rate $\gamma$. If we denote the recovery time of the $i$-th individual by $T_{i,R}$, it follows immediately from \Cref{eq:micro_RTCM_stoch_equations} that $T_{i,R}-T_{i,I}$ and $T_{i,I}$ are independent and $T_{i,R}-T_{i,I}$  follows an exponential distributions with rate $\gamma$. Symbolically,
\begin{align}
 T_{i,R}-T_{i,I} \perp  T_{i,I}  \text{ and }  T_{i,R}-T_{i,I} \sim \sys{Exponential}(\gamma) \eqstop  \label{eq:Sellke_independence}
\end{align}
Note that the fate of an individual is entirely described by the statistical distributions given in \Cref{eq:infection_time_distribution,eq:Sellke_independence}. 
It is also interesting to note that an individual's fate depends on the  process history ${\cal H}_t$ only through the variable $\beta\int_0^t I(s)\, \differential{s}/n$ akin to an improper  cumulative hazard function (improper,  since  $ \int_0^\infty I(u) \differential{u}<\infty$ with probability one). These considerations lead to \Cref{alg:Sellke} for simulating the process in  \Cref{eq:micro_RTCM_stoch_equations}. This is known  as the \emph{Sellke construction} \cite{britton2000stochEpi,fleming2005countingprocess,aalen2008survival} in the literature.   It can be easily verified that \Cref{alg:Sellke} is equivalent to simulating the system in \Cref{eq:micro_RTCM_stoch_equations} using the Doob-Gillespie algorithm.  As we describe below, the Sellke construction  plays a central role in developing survival representations of the \ac{SIR} \Cref{eq:SIR_ODE}. Moreover, it also turns out  to be  equivalent to a statistical representation of micro models under the law of mass action based on {\em contact intervals} \cite{kenah2010contact,kenah2013nonparametric}.

\begin{algorithm}[ht]
\small\topsep=0in\itemsep=0in\parsep=0in
  \begin{algorithmic}[1]
    \caption{\label[algorithm]{alg:Sellke}%
  \small   Pseudocode for the Sellke construction}
\State     Assume you have initially $m$  infectives and $n$ susceptibles. Arrange all $n$  susceptibles according to the  order statistics $Q_{(1)}<\ldots<Q_{(n)}$ of an \ac{iid} random sample from $\sys{Exponential}(1)$
\State Simulate $m+n$ infectious  periods as \ac{iid}  sample from $\sys{Exponential}(\gamma)$
\State Set $i=1$
\State Calculate $\Lambda(t)= \frac{\beta}{n} \int_0^t I(u) \differential{u}$ and update it with removal  times from Step 2
\State Calculate $t_i=\inf \{ t:  Q_{(i)}>\Lambda(t)\}$.  If $t_i<\infty$ change   $i$-th  susceptible to infective,  update $\Lambda(t_i)$, else Stop
\State Set $i=i+1$ and $t=t_i$ and go to Step 4.
  \end{algorithmic}
\end{algorithm}


\subsection{Mean-field limit of \ac{SIR}}
The simplest way to derive a macro model from the micro description 
is via lumping or aggregation of states. When the aggregation of states is  \emph{strongly lumpable} \cite{kemeny1983finite,rubino_1989,RUBINO1993WL_CTMC,buchholz_1994}, the resultant aggregated process remains Markovian for any choice of the initial distribution. Now, for the \ac{SIR} process, let $\mathcal{X}\defeq \{\S,\I,\R\}$ denote the possible statuses of the individuals. Then, $\mathcal{X}^{n+m}$ is the state space of the ensemble of individual-based $S_i, I_i, R_i$ processes. Define the macro-level processes
\begin{align}
  S(t)=\sum_{i = 1}^{n+m} S_i(t),   I(t)=\sum_{i = 1}^{n+m} I_i(t), \text{ and }   R(t)=\sum_{i = 1}^{n+m} R_i(t) \eqcomma
\end{align}
which keep track of the total counts of susceptible, infected and recovered individuals. Let $L \defeq \binom{n + m + 2}{2}$. Partition $\mathcal{X}^{n+m}$ into $\mathcal{X}_1, \mathcal{X}_2, \ldots, \mathcal{X}_L$ such that any two states in each $\mathcal{X}_l$ produce the same counts for $S(t), I(t), R(t)$, for $l=1,2,\ldots, L$. It is easy to see that the Markov chain described by the ensemble in \Cref{eq:micro_RTCM_stoch_equations} is (strongly) lumpable with respect to the partition $\{\mathcal{X}_1, \mathcal{X}_2, \ldots, \mathcal{X}_L\}$ (see \cite{kemeny1983finite,khudabukhsh2018Lumpability,Simon2011Exact}). That is, the lumped process $(S,I,R)$ is also Markovian for any choice of the initial distribution. Therefore, we can write
\begin{align}
  \begin{aligned}
  S(t) ={}& S(0) - Y\Big( \int_0^t \frac{\beta}{n}  S(s) I(s) \, \differential{s}  \Big) \eqcomma \\
  I(t) ={}& I(0) + Y \Big( \int_0^t \frac{\beta}{n}  S(s) I(s) \, \differential{s}  \Big) - Z \Big( \int_0^t \gamma I(s)\, \differential{s}\Big) \eqcomma \\
  R(t) ={} & Z\Big( \int_0^t \gamma I(s)\, \differential{s}\Big) \eqcomma
  \end{aligned} \label{eq:macro_SIR_RTCM_system}
\end{align}
where $Y$ and $Z$ are independent unit rate Poisson processes. As before, the simulation of the above system can be done using the Doob-Gillespie algorithm. For the sake of completeness, we present a pseudocode   in \Cref{alg:Gillespie}.
\begin{algorithm}[ht]
\small\topsep=0in\itemsep=0in\parsep=0in
\begin{algorithmic}[1]
  \caption{\label[algorithm]{alg:Gillespie}%
  \small Pseudocode for Doob-Gillespie algorithm}
  \State Initiate $(S(0),I(0),R(0))$
  \State Assume you have the process value $(S(t), I(t),R(t))$ at $t\ge 0$
  \State Calculate rates $\lambda_I(t)=\beta S(t)I(t)/n$  and $\lambda_R(t)=\gamma\, I(t)$
  \State Set   next  transition  time  $\Delta t$  as $\sys{Exponential}(\lambda_I(t)+\lambda_R(t))$
\State   Select  transition type  (infection or recovery) as $\sys{Bernoulli}\left( \frac{\lambda_I(t)}{\lambda_I(t)+\lambda_R(t)}\right)$
  \State Update $(S(t'), I(t'),R(t'))$ at $t'= t+\Delta t$ and go to Step 2. 
\end{algorithmic}
\end{algorithm}
The macro model is particularly convenient in that it is amenable to asymptotic analysis. Indeed, for very large populations, we can approximate the stochastic \ac{SIR} dynamics by a system of \acp{ODE}. The rationale behind this approximation is that pure jump Markov processes approach solutions of a certain \ac{ODE} in the limit, when scaled appropriately  \cite{kurtz1970solutions,kurtz1978strong}. This is sometimes called \emph{mean-field} or \emph{fluid limit} of the Markov jump process.


For our \ac{SIR} system in \Cref{eq:macro_SIR_RTCM_system}, the scaled process $({S}_n,{I}_n,{R}_n) \defeq (\frac{S}{n}, \frac{I}{n},\frac{R}{n}   ) $ satisfies
\begin{align}
  \begin{aligned}
{S}_n(t)={}& {S}_n(0) - n^{-1}    Y\Big( n \int_0^t \beta  {S}_n(s) {I}_n(s) \, \differential{s}  \Big) \eqcomma \\
{I}_n(t) ={}& {I}_n(0) + n^{-1}   Y \Big( n \int_0^t {\beta} {S}_n(s) {I}_n(s) \, \differential{s}  \Big) - n^{-1}  Z \Big( n \int_0^t \gamma {I}_n(s)\, \differential{s}\Big) \eqcomma \\
{R}_n(t) ={} & n^{-1}   Z\Big( n \int_0^t \gamma {I}_n(s)\, \differential{s}\Big) \eqstop
  \end{aligned}
  \label{eq:scaled_stoch_SIR_RTCM}
\end{align}
By virtue of the Poisson \ac{LLN} \cite{ethier2009markov}, which asserts that $ n^{-1} V(nt)\approx t $ for large $n$ and a unit rate Poisson process $V$, we see that the processes in \Cref{eq:scaled_stoch_SIR_RTCM} converge to the
solution of the following system of \acp{ODE} as $n\rightarrow \infty$ and $m/n \rightarrow \rho \in (0,1)$:
\begin{align}
\begin{aligned}
\dot{s_t} ={} & -\beta s_t \iota_t \eqcomma \quad
\dot{\iota_t} ={}  \beta s_t \iota_t - \gamma \iota_t \eqcomma \quad
\dot{r_t} ={} & \gamma \iota_t \eqcomma
\end{aligned} \label{eq:mean_field_SIR}
\end{align}
which are the same as the Kermack and McKendrick \acp{ODE} in \Cref{eq:SIR_ODE}.  The introduction of $\rho$ is convenient because it sets $s_0=1$, $\iota_0=\rho$ and $r_0=0$. The rate of convergence to this \ac{LLN} \ac{ODE} limit can be computed using sample path \ac{LDP} of the Markov process in \Cref{eq:scaled_stoch_SIR_RTCM}. Standard tools from \cite{feng2006large,dembo2010LargeDeviations,Dupius2011LDP} as well as related results from \cite{dolgoarshinnykh2009sample,pardoux2017large,djehiche1998large} can be borrowed for this purpose. However, our main motivation here is to interpret \Cref{eq:mean_field_SIR} as describing an \ac{MTM}. We make this point precise in the following.

\setcounter{equation}{0}
\section{\acl{SDS}}
\label{sec:mtm}
The \acp{ODE} in  \Cref{eq:mean_field_SIR} describing the mean-field macro model can be given a probabilistic interpretation. It is convenient to rewrite   \Cref{eq:mean_field_SIR} as follows:
\begin{align}
 \begin{aligned}
s_t ={} &  \myExp{ -\beta \int_0^t \iota_u \, \differential{u}}  = \myExp{-\ro r_t } \eqcomma \\
\iota_t = {}&  \rho e^{-\gamma t} - \int_0^t \dot{s_u} e^{-\gamma (t-u)  } \, \differential{u} \eqcomma \\
r_t ={} &  \gamma \int_0^t \iota_u \, \differential{u} \eqcomma
 \end{aligned}
\label{eq:mean_field_SIR_integral_form}
\end{align}
where  $\ro=\frac{\beta}{\gamma}$ is the basic reproduction number. Here,  the first two equations are obtained by partially solving the \ac{ODE} system using the integrating factor (first equation) and the variation of parameter method (second equation).

\begin{figure}[t]
\centering
  \begin{tikzpicture}
  \begin{scope}
\node(box1)[draw,rectangle, minimum size=1cm,line width=1pt] at (1,1){Susceptible (S)};
\node(box2)[draw,rectangle, minimum size=1cm,line width=1pt] at (6.5,-1){Infected (I)};
\node(box3)[draw,rectangle, minimum size=1cm, line width=1pt] at (13.0,-1){Recovered (R)};
\node(box4)[draw,rectangle, minimum size=1cm, line width=1pt] at (6.5,3){Never infected};
  \draw  [-latex, line width=1pt](box1) to node [minimum size=0.3cm, xshift=2.5cm, yshift=0.30cm] {with probability~$\tau$: $T_{I} \sim f_{\tau}(t)$ } (box2);
  \draw  [-latex, line width=1pt,in=180, out=0](box2) to node [minimum size=0.3cm, xshift=0.1cm, yshift=0.30cm] {$T_{R}\sim g_{\tau}(t)$} (box3);
 \draw  [-latex, line width=1pt](box1) to node [minimum size=0.3cm, xshift=2.5cm, yshift=-0.10cm] {with probability~$1-\tau$} (box4);
  \end{scope}
    \begin{pgfonlayer}{background}
    \filldraw [line width=4mm,join=round,black!05]
      (box4.north  -| box1.west)  rectangle (box3.south  -| box3.east)
      (box4.north -| box1.west) rectangle (box3.south -| box3.east);
  \end{pgfonlayer}
  \end{tikzpicture}
    \caption[\ac{SIR} as SDS]{\label{fig:transfer}%
\small \ac{MTM}/\ac{SDS} derived from \ac{SIR} \Cref{eq:SIR_ODE}. To each individual, we assign random variables $T_I$ and $T_R$ specifying his/her transfer times. The laws of $T_I$ and $T_R$ are given by \Cref{eq:f,eq:g}.}
\end{figure}
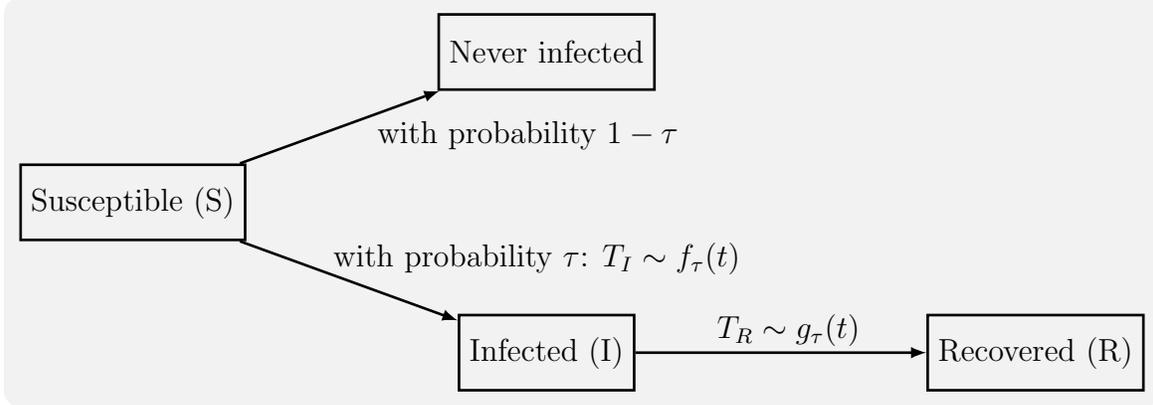

 We will now interpret \Cref{eq:mean_field_SIR_integral_form}  as describing the  mass (probability)  transfer model  in an infinite population where a randomly selected  unit transfers over time from the initial susceptible (S) compartment first to the infected   (I) and then to the removed (R)  compartment. This  idea is depicted in \Cref{fig:transfer}. 
The transfer process is described by the  state (compartment allocation) of a randomly selected  unit (say, $U_t$)  at $t$ where  $U_t\in \{\S,\I,\R\}$ with $U_0=\S$.


According to the mass transfer  interpretation of the  \Cref{eq:mean_field_SIR_integral_form},  the time of infection (transfer from \S\ to \I) of $U_t$ is  given by the (improper) random variable $T_I=\inf\{t>0: U_t=\I\}$  with its distribution determined by the function
$s_t=\myExp{-\ro r_t}$, that is
\be{eq:agg_surv1} \probOf{T_I>t}=s_t.
\ee
Note that this is a direct analogue of \Cref{eq:infection_time_distribution} in our  aggregated macro model where   the stochastic quantity $\int_0^t I(u)\, \differential{u}/n$ is replaced by its  deterministic limit  $\int_0^t \iota_u\, \differential{u}$ from  \Cref{eq:mean_field_SIR_integral_form}.  It is important to note that in the limit, the units become independent. This phenomenon is also known in the literature as \emph{mean-field independence} or \emph{propagation of chaos} \cite{mcdonald2007lecture,baladron2012mean,meleard1996asymptotic}.

None also  that although by our assumption   $s_0=1$ and  $s_t$ is non-increasing, it is nevertheless an {\em improper survival function} since $P(T_I=\infty)=  s_\infty=1-\tau>0$  where
  $\tau=r_\infty-\rho$ satisfies the deterministic final size equation
  \begin{align}
  1-\tau=\myExp{-\ro (\tau+\rho)} \eqcomma
  \label{eq:tau}
  \end{align}
  which is a contraction map and therefore, numerically amenable to efficient fixed-point iteration schemes.
Since $0<\tau<1$,  we may interpret $\tau$ as   the probability of unit $u$ ever transferring  out of  the compartment $\S$ (ever being infected). Consequently,
 $\ro r_t=\beta \int_0^t \iota_s \differential{s}$  may be thought of as   the (improper)  {\em cumulative hazard} for $u$ and
$\beta \iota_t$ as  the (improper) {\em hazard function} of the improper random variable $T_I$. This hazard is sometimes called the \emph{force of infection}.
  By the law  of total probability $P(T_I>t)=s_t=\tau \tilde{s}_t+1-\tau $.
 Here $\tilde{s}_t=(s_t-1+\tau)/\tau$ is the proper conditional survival function,
  conditioned on $T_I<\infty$,  that is,  on an event that individual $u$  ever gets  infected  (\ie,  transfers  out of   $\S$).
Note that according  to \Cref{eq:mean_field_SIR_integral_form} the density for $\tilde{s}_t$   is

\begin{align}
f_{\tau}(t)=\frac{\beta}{\tau} s_t \iota_t =-\frac{\dot{s}_t}{\tau} \quad \text{ for }  t>0 \eqstop
\label{eq:f}
\end{align}


Let $T_R=\inf \{t>0: U_t=\I\}$ be the removal  time  of a  unit $u$ with exposure time $T_I$. Note that from  \Cref{eq:mean_field_SIR_integral_form} we obtain   using  \eqref{eq:f}
\be{eq:conv}
\frac{\ga\tilde{\iota}_t}{\tau} = \int_0^t f_\tau(u)\, \ga e^{-\ga(t-u)} \differential{u} \eqcomma  \\
\ee
where ${\tilde \iota}_t =\iota_t-\rho \myExp{-\ga t}$. Since   $f_\tau(u)$ is a density function,  the right hand side above is a convolution of the density of $T_I$ and the (exponential) density of $T_R-T_I$. It thus follows that the right hand side quantity
\begin{align}
g_{\tau}(t)= \gamma (\iota_t- \rho e^{-\gamma t})/\tau
\label{eq:g}
\end{align}
 is itself a density of  the variable $T_R$, which is  the sum of two independent random variables
$T_I$ and $T_R-T_I$ (that is $T_R \ \bot \ T_R-T_I$). Note the analogy of this  result with \Cref{eq:Sellke_independence}.
These considerations  give us \Cref{alg:massTransfer} for simulating the individual histories in the \ac{MTM} \ac{SIR} model. See also \Cref{fig:transfer} for a pictorial representation of the idea.

\begin{algorithm}[H]
\small\topsep=0in\itemsep=0in\parsep=0in
\begin{algorithmic}[1]
\caption{\label[algorithm]{alg:massTransfer}
\small Pseudocode for simulating a single trajetory from \ac{MTM}
}
\State Calculate  $(s_t, \iota_t, r_t)$ as given  by  \Cref{eq:mean_field_SIR}
\State With probability $1-\tau$,  where $\tau$ is given by \Cref{eq:tau},  leave the unit  in $S$ state forever. With probability $\tau$  move to Step 3
\State Simulate infection time $T_I \sim f_\tau(t)$ where the density $f_\tau(t)$  is given by \Cref{eq:f}
\State Independent  of $T_I$, simulate  infectious period $T_R-T_I\sim \sys{Exponential}(\gamma)$
\State Record  the pair $(T_I, T_R)$.
\end{algorithmic}
\end{algorithm}

\begin{myRemark}
   \label{MTM_solutions}
  Note that analyzing  timepoints  $(T_I,T_R)$ according to the  \Cref{alg:massTransfer}  addresses  all four issues of macro \ac{SIR} model in \Cref{eq:SIR_ODE} described in \Cref{sec:intro}.  Indeed, \Cref{alg:massTransfer} no longer requires the population size (problem~\ref{problem1}).  Direct generation of individual trajectories  according to  \Cref{alg:massTransfer} also  allows us to specify a likelihood function (problem~\ref{problem4}), account for differences in individual characteristics (problem~\ref{problem2}), and overcome issues with censoring or interval-based data (problem~\ref{problem3}).
 \end{myRemark}

 In a way,  \Cref{alg:massTransfer} brings us back from the macro to  the micro  level and completes  the conceptual ``micro-macro-micro'' loop. The mass transfer interpretation has similarities with \emph{symbolic dynamical systems} \cite{hao1989elementary,lind1995introduction,kakutani1972strictly}.

\setcounter{equation}{0}
\section{Parameter inference}
\label{sec:inference}
Under the stochastic (agent-based) micro \ac{SIR} model \Cref{eq:micro_RTCM_stoch_equations} or its aggregated macro  version in \Cref{eq:macro_SIR_RTCM_system},  the  vector of parameters of interest is  $\theta=(\beta,\ga,\rho)$ with $m=I(0)=\rho n$, since the  parameter $\tau$ is expressible in terms of $\theta$ via \Cref{eq:tau}.  The size of the initial  susceptible population  ($n$) is   usually unknown and may be considered  a nuisance parameter. The estimation of this nuisance parameter is often problematic, and popular methods such as profile likelihoods do not always yield good estimates. In order to address this problem,  we propose  the \ac{SDS} likelihood, which is based on the \ac{SDS} interpretation of the \ac{SIR} \Cref{eq:SIR_ODE}  and does not require $n$. Before going into the details of \ac{SDS} likelihood, we describe the exact likelihood based on the Doob-Gillespie \Cref{alg:Gillespie}.   To emphasize the strength of our \ac{SDS} likelihood and compare its performance against the exact likelihood,  we assume that the value of $n$ is available for the exact likelihood.

\subsection{Exact (Doob-Gillespie) likelihood}
\label{sec:inference:gillespie}
Assume that there were total of $z=z_I+z_R$ events $(k_i,t_i)_{i=0}^z$ up to time $T$ of which $z_I$ are infections and $z_R$ are removals at  times $0<t_1<\ldots <t_z=T$,  where $k_i\in\{\I,\R\}$ denotes the type of the event.  Put $X(t)= (S(t),I(t),R(t))$. Then, following \Cref{alg:Gillespie}, the exact  log-likelihood for $\theta$ is
\begin{align}
\label{eq:lkh1}
\ell_1(\theta\mid X(t)_{t\in [0,T]}) &= \sum_{i=1}^z \log\left(  \lambda_{k_i}(X(t_i)) \right)- \int_0^T [\lambda_I(X(t))+\lambda_R(X(t))]\, \differential{t} \nonumber \\
& =z_I \log\left( \beta \right) +z_R \log \left( \gamma \right) + \sum_{i: k_i=\I} \log \left( S(t_i)/n \right)  \nonumber\\ &{}\quad +\sum_{i=1}^z \log \left( I(t_i) \right) - \int_0^T \frac{ \beta}{n} S(t)I(t)\,  \differential{t} -\int_0^T \gamma I(t)\, \differential{t} \eqcomma
\end{align}
where the last two integrals may be also written as finite sums.  From \Cref{eq:lkh1} the \acp{MLE}  for  $\beta$ and $\gamma$ can be derived as
\be{eq:mle1} \hat{\beta}= \frac{n z_I}{\int_0^T  S(t)I(t) \, \differential{t}} \quad\text{and}\quad  \hat{\gamma}= \frac{z_R}{\int_0^T I(t) \, \differential{t}}\eqstop    \ee
Because we assume we know the population size $n$ and the trajectory $X(t)_{t \in [0, T]}$ for the exact likelihood, the parameter $\rho= I(0)/n$ is known exactly.


\subsection{\ac{SDS} likelihood}
\label{sec:inference:mtm_likelihood}
Following the discussion in \Cref{sec:mtm}, an approximation of the exact likelihood function $\ell_1(\theta)$ in \Cref{eq:lkh1} can be obtained from \Cref{eq:infection_time_distribution} by replacing the process $I(u)/n$ with its limit $\iota$ (as $n\to \infty$) and considering the individual trajectories as independent. Since we let $n\to \infty$, the exact value of the initial size of the susceptible population is no longer needed.

Assume we randomly sample $N+M$ individuals of whom $N$ are found susceptible and $M$, infected. We observe those $N+M$ individuals up to the cut-off time $T$ and record their infection or recovery times. Suppose $K$ out of the $N$ initially susceptible individuals get infected  at infection times~$t_1,t_2,\ldots,t_K$ and $L$ of them recover by time~$T$. Pair each infection time~$t_i$ with the corresponding duration of infectious period $\omega_i$ if the individual recovers by time~$T$. If the individual does not recover by time $T$, pair $t_i$ with the censored information~$\omega_i=T-t_i$. Among the $M$ initially infected individuals, suppose $\tilde{L}$ individuals recover by the cutoff~$T$ at times $\epsilon_1, \epsilon_2,\ldots, \epsilon_{\tilde{L}}$. Then, following \Cref{alg:massTransfer}, we have the following \ac{SDS} likelihood
\begin{align}
\label{eq:lkh2}
\ell_2(\theta \mid \{t_i,\omega_i\}_{i=1}^{K}, \{\epsilon_j\}_{j =1}^{\tilde{L}} ) ={}&  (N-K)\log\left( s_{T} \right)+ \sum_{i=1}^{K} \log \left( \tau f_{\tau}(t_i) \right) + (L+\tilde{L})\log\left( \gamma \right)    \nonumber \\
& {} -\gamma \left( \sum_{i=1}^{K} \log\left( \omega_i \right) +  \sum_{j=1}^{\tilde{L}} \log\left( \epsilon_j \right) + (M-\tilde{L})T \right) \eqcomma
\end{align}
where, as described in \Cref{sec:mtm},
\begin{align*}
f_\tau(t) &= {\beta}{\tau}^{-1} \iota_t \myExp{-\ro t_t} \eqcomma \quad
s_t =  \myExp{-\ro r_t} \eqcomma
\end{align*}
and  $\tau=r_\infty-\rho$ satisfies  \Cref{eq:tau}.  Performance of the \ac{SDS} likelihood given in \Cref{eq:lkh2} and \ac{MCMC} implementations are  discussed in the next section.

%
%

\setcounter{equation}{0}
\section{Numerical examples}
\label{sec:numerical}
\subsection{Bayesian estimation using \ac{MCMC}}
\label{subsec:sds_mcmc}
In this section, we present numerical examples to illustrate how one can use the \ac{SDS}-likelihood in~\Cref{eq:lkh2} to infer the unknown parameter $\theta$ using \ac{MCMC} methods.
In order to construct a posterior distribution for $\theta$, we assign gamma priors to the  parameters $\beta$,  $\gamma$ and  $\rho$: 
\begin{align}
\begin{aligned}
\beta \sim& ~\sys{Gamma}(a_{\beta}, b_{\beta}) \eqcomma  \\
\gamma \sim& ~\sys{Gamma}(a_{\gamma}, b_{\gamma}) \eqcomma  \\
\rho \sim& ~\sys{Gamma}(a_{\rho},b_{\rho}) \eqstop
\end{aligned}
\label{eq:priors}
\end{align}
The positive quantities $a_{\beta}, b_{\beta}, a_{\gamma},b_{\gamma}, a_{\rho}$, and $b_{\rho}$ are appropriately chosen hyper-parameters. The posterior distribution of $\theta$ is obtained by Bayes' rule: It is proportional to the product of the likelihood function given in \Cref{eq:lkh2} and above three priors. However, the posterior distribution cannot be written in closed-form. Even if a conditional posterior distribution is obtained, we can not find any closed-form expression for the probability density function because we need to have solutions $s_{t}$, $\iota_{t}$, $r_{t}$ to \Cref{eq:mean_field_SIR_integral_form}, which are also functions of $\theta$. Thus, we cannot immediately employ  a generic Gibbs sampler method \cite{smith1993bayesian,choi2011inference}.  Therefore,  we need a more efficient  updating algorithm than the standard Metropolis-Hastings algorithm. In this paper, we adopt  the \ac{RAM} algorithm\footnote{The \ac{RAM} method generalizes the \ac{ASM} algorithm \cite{atchade2005adaptive} by updating the tuning constant appropriately  to achieve  optimal acceptance ratio.}  \cite{vihola2012robust, mcrae2014bayesian}, which adjusts the tuning constant and the variance-covariance matrix of the proposal distribution adaptively  to maintain a high acceptance ratio in the Metropolis steps. The variance-covariance matrix is updated during the \ac{MCMC} iterations.



\begin{algorithm}[ht]
	\small\topsep=0in\itemsep=0in\parsep=0in
	\begin{algorithmic}[1]
		\caption{\label[algorithm]{alg:MCMC}
			\small  \ac{MCMC} for drawing posterior sample using \ac{RAM} method
		}
		\State Initialize $(\beta, \gamma, \rho)$ and the variance-covariance matrix of proposal distribution
		\Repeat
		\Comment{adjust for burn-in etc.}
		\State Draw candidate samples of $(\beta, \gamma, \rho)$ from the proposal distribution
		\State Solve  \Cref{eq:mean_field_SIR} and store the solutions at the observed infection times $t_1, t_2, \ldots t_K$
		\State Run Metropolis algorithm and determine whether the candidate samples are accepted
		\State Run \ac{RAM} method to update the variance-covariance matrix of the proposal distribution
		\Until{convergence.}
	\end{algorithmic}
\end{algorithm}

\subsection{Simulation study}
\label{subsec:simulation_study}
In order to compare the accuracy of the inference based on the \ac{SDS}-likelihood  against the exact (Doob-Gillespie) likelihood, we performed simulation studies under various sets of parameters and size of the susceptible population. We also consider the impact of various truncation times. The data used for parameter inference are generated according to \Cref{alg:Sellke}. 

We compare four different inference methods. We list them below:
\begin{enumerate}
\item \textbf{Method 1} The first method uses the Doob-Gillespie likelihood given in \Cref{eq:lkh1}  and calculates \acp{MLE} according to  \Cref{eq:mle1}.
\item \textbf{Method 2} The second method also uses the Doob-Gillespie  likelihood given in \Cref{eq:lkh1}, but implements an \ac{MCMC} scheme with the priors listed in \Cref{eq:priors}  to infer $\theta$. Because of conjugacy of the gamma priors, the posteriors are also gamma distributions \cite{choi2011inference}. In particular, they are given by
\begin{align*}
{\beta \mid (X(t)_{t\in [0,T]} }) \sim &~\sys{Gamma}(nz_I + a_{\beta}, \int_0 ^T S(t)I(t)\differential{t} + b_{\beta})  \eqcomma \\
\gamma \mid (X(t)_{t\in [0,T]} ) \sim &~\sys{Gamma}(z_R + a_{\gamma}, \int_0 ^T I(t)\differential{t} + b_{\gamma}).
\end{align*}
\item \textbf{Method 3} The third method uses the \ac{SDS}-likelihood given in \Cref{eq:lkh2} and follows the \ac{MCMC} procedure described earlier in \Cref{alg:MCMC}.
\item \textbf{Method 4} When $n$ is large, one can perform a diffusion approximation of the process \eqref{eq:macro_SIR_RTCM_system} and replace the  likelihood in \Cref{eq:lkh1} with the Gaussian likelihood.  Our fourth method uses the Gaussian likelihood  in \Cref{eq:lkh3} for implementing  an \ac{MCMC} scheme. See \Cref{appendix:gaussian} for more details.
As the likelihood of $\beta$ and $\gamma$ has the Gaussian form, we  assign conjugate normal prior to $\beta$ and $\gamma$ with same mean and variance as the gamma priors mentioned in  \Cref{eq:priors}, \ie,
\begin{align*}
\beta \sim ~ 	 \sys{N} \left(\frac{a_{\beta}}{b_{\beta}},\frac{a_{\beta}}{b_{\beta}^2} \right) \eqcomma \quad \text{ and } \quad
\gamma \sim  ~ 	 \sys{N} \left(\frac{a_{\gamma}}{b_{\gamma}},\frac{a_{\gamma}}{b_{\gamma}^2} \right) \eqstop
\end{align*}
The conditional posterior distributions of $\beta$, and  $\gamma$ satisfy
\begin{align*}
\beta\mid (X(t)_{t\in (0,T]},\gamma, \rho)  &~\sim \sys{N}\bigg(\mu_{\beta}
,	\sigma_{\beta}^2\bigg) \eqcomma  \\
\gamma \mid (X(t)_{t\in (0,T]},\beta, \rho ) &~\sim \sys{N}\bigg(\mu_{\gamma}
,	\sigma_{\gamma}^2\bigg)  \eqcomma  \\
\end{align*}
where
\begin{align*}
\mu_{\beta} &{} = \left(\frac{z_I}{(1-s_T)}+ b_\beta\right)\left(\frac{\int_0 ^T S(t)I(t)\differential{t}}{n(1-s_T)}+\frac{b_{\beta}^2}{a_\beta}\right)^{-1} \eqcomma \\
\sigma_{\beta}^2 &{} = \left(\frac{\int_0 ^T S(t)I(t)\differential{t}}{n(1-s_T)}+\frac{b_{\beta}^2}{a_\beta}\right)^{-1} \eqcomma \\
\mu_{\gamma} &{} = \left(\frac{z_R}{n(1+\rho-s_T-\iota_T)}+ b_\gamma\right)\left(\frac{\int_0 ^T I(t)\differential{t}}{ n(1+\rho-s_T-\iota_T)}+\frac{b_{\gamma}^2}{a_\gamma}\right)^{-1} \eqcomma \\
\sigma_{\gamma}^2 &{} = \left(\frac{\int_0 ^T I(t)\differential{t}}{ n(1+\rho-s_T-\iota_T)}+\frac{b_{\gamma}^2}{a_\gamma}\right)^{-1} \eqstop
\end{align*}
However, the conditional posterior distribution of $\rho$, which is given by
\begin{align*}
\rho \mid (X(t)_{t\in (0,T]},\beta, \gamma)  &~\propto \left(2\pi(1-s_T)(1+\rho-s_T-\iota_T) \right)^{-1} \times  \sys{Gamma}(a_{\rho},b_{\rho}) \eqstop
\end{align*}
does not assume a simplified form.
 Note that with  independent priors,  the   posterior distributions of $\beta$ and $\gamma$  are  independent conditionally on $\rho$ and that  the conditional  posterior distribution of  $\rho$  depends only on the prior parameters and  the  solution of \Cref{eq:mean_field_SIR}. 
 However, in order to draw posterior samples of $\rho$ conditional on $\beta$ and $\gamma$, we need to apply the  Metropolis algorithm  \cite{tierney1994markov}.
\end{enumerate}

For all \ac{MCMC}-based methods, we add  constraints on the proposed values of $\rho$  in the \ac{MCMC} iteration steps so that $\rho$ remains within $(0,1)$ and satisfies \Cref{eq:tau}. We have a total of 18 simulation scenarios based on the combinations of the following:
\begin{itemize}
  \item Three values of $\theta=(\beta, \gamma, \rho)$ : $\theta_1 =  (2.0, 0.5, 0.05)$, $\theta_2 = (2.0, 1.0, 0.05)$, and $\theta_3 = (1.5, 1.0, 0.05)$ yielding  the basic reproduction number $\ro$ equal to 4, 2, and 1.5 respectively.
  \item Two cutoff times $T$: One cutoff time is chosen around  the half-time  of the epidemic duration (at the peak of the infection process) and another one  towards the end. Therefore, the chosen values of $T$ are 3 and 9 for $\theta_1$, 3 and 7 for $\theta_2$, and  3 and 6 for $\theta_3$. See \Cref{fig:3SIR} in \Cref{appendix:inference} for the \ac{SIR} curves for different parameter values and cutoff times. The vertical line in the plot represents the cutoff time.
  \item Three values of the size of the susceptible population $n$: $10^2, 10^3$, and $10^4$.
\end{itemize}

For each of the 18 scenarios, we generate $10^2$ sets of synthetic epidemic data using  \Cref{alg:Sellke}. Each  generated data set has $n + n\times\rho$ rows and two columns. Each row corresponds to an individual in the epidemic and two columns are the individual's infection  $T_I$ and removal times $T_R$. In order to ensure the prior distributions in our Bayesian inference are not too informative, we set $a_{i} = i \times 0.01$ and $b_{i} = 0.01$ for $i = \beta, \gamma$, and $\rho$.  We iterated the \ac{MCMC} procedures 11,000 times  for both {Method 3} and {Method 4}. The first 1,000 iterations are removed as burn-in. After burn-in, every 10th iteration is stored as a posterior sample. In total, 1,000 posterior samples are used for estimation. For {Method 2}, we generate 1,000 samples without any burn-in phase or thinning because Monte Carlo simulations are sufficient. For the Bayesian methods (\ie, {Method 2}, {Method 3}, and {Method 4}), we estimate the parameters $\beta$, $\gamma$, and $\rho$ by taking the means of 1,000 posterior samples.

Since $\rho$ is a $(0,1)$-valued random variable, assigning  $\sys{Beta}(1,1)$ as prior is natural. However, in this simulation study, we assign a slightly more informative gamma prior $\sys{Gamma}(\rho\times 0.1, 0.1)$. The reason for this choice is that  the conditional posterior of $\rho$ depends only on the solution of   \Cref{eq:mean_field_SIR} when we implement the \ac{MCMC} procedure for  {Method 4} using a Gaussian likelihood. As a consequence, the estimates for $\rho$ using Method~4 are very poor and with $\rho$ prior $\sys{Beta}(1,1)$, the posterior means are always around $0.5$, the prior mean. In order to circumvent this limitation of {Method 4}, we assign a $\sys{Gamma}(\rho\times 0.1, 0.1)$ prior, which is slightly more informative. It is important to note  that the performance of our Method~3 based on the \ac{SDS}-likelihood remains unaffected even if an uninformative $\sys{Beta}(1,1)$ prior is chosen for $\rho$. The choice of a $\sys{Gamma}(\rho\times 0.1, 0.1)$ prior for $\rho$ is made only to ensure fair comparisons of the different inference methods.


\begin{sidewaystable}[ph!]
		\footnotesize
		\centering
		\begin{tabular}{|c|c|c|c|c|c|c|c|c|c|c|c|c|}
			\hline
			& \multirow{2}{*}{ n } & \multirow{2}{*}{ Statistics } & \multicolumn{4}{|c|}{ $\beta$ } & \multicolumn{4}{|c|}{ $\gamma$ } & \multicolumn{2}{|c|}{$\rho$} \\
			\cline{4-13}
			& & & Method~1 & Method~2 & Method~3 & Method~4 & Method~1 & Method~2 & Method~3 & Method~4 & Method~3 & Method~4\\
			\hline
			& \multirow{2}{*}{ $10^4$ } & Avg. & 2.00666 & 2.00666 & \myhighlight{2.00164} & 2.00673 & \myhighlight{0.50024} & 0.50025 & 0.49928 & 0.50027 & \myhighlight{0.04984} & 0.1771 \\
			{$\beta$=2 }& & (MSE) & (0.00046) & (0.00046) & (0.00062) & (0.00045) & (0.00002) & (0.00002) & (0.00002) & (0.00002) & (0.00001) & (0.01864) \\
			\cline{2-13}
			\multirow{2}{*}{ $\gamma$=0.5 }& \multirow{2}{*}{ $10^3$ } & Avg. & 2.00326 & 2.00326 & 2.00465 & \myhighlight{2.00325} & 0.49963 & \myhighlight{0.49965} & 0.49788 & 0.49958 & \myhighlight{0.04968} & 0.17308 \\
			& & (MSE) & (0.00334) & (0.00334) & (0.00897) & (0.00336) & (0.00024) & (0.00024) & (0.00028) & (0.00024) & (0.00013) & (0.01775) \\
			\cline{2-13}
			{ $\rho$=0.05 }& \multirow{2}{*}{ $10^2$ } & Avg. & 2.04332 & 2.04317 & \myhighlight{2.02667} & 2.04421 & 0.50553 & 0.50553 & 0.48972 & \myhighlight{0.50542} & \myhighlight{0.05578} & 0.20759 \\
			& & (MSE) & (0.04238) & (0.04236) & (0.07655) & (0.04303) & (0.00284) & (0.00282) & (0.00278) & (0.00283) & (0.00160) & (0.02583) \\
			\cline{2-13}
			\hline
			& \multirow{2}{*}{ $10^4$ } & Avg. & 2.00257 & \myhighlight{2.00065} & 2.00137 & 2.00258 & 1.00033 & \myhighlight{0.99991} & 1.00057 & {1.00029} & \myhighlight{0.04982} & 0.19275 \\
			{$\beta$=2 }& & (MSE) & (0.00046) & (0.00044) & (0.00101) & (0.00046) & (0.00012) & (0.00013) & (0.00016) & (0.00012) & (0.00001) & (0.02311) \\
			\cline{2-13}
			\multirow{2}{*}{ $\gamma$=1 }& \multirow{2}{*}{ $10^3$ } & Avg. & 1.99424 & 1.99424 & 1.98761 & \myhighlight{1.99434} & 0.99613 & 0.99595 & 0.99199 & \myhighlight{0.99614} & \myhighlight{0.04963} & 0.19336 \\
			& & (MSE) & (0.00489) & (0.00489) & (0.01062) & (0.0049) & (0.00107) & (0.00108) & (0.00132) & (0.00107) & (0.00016) & (0.02405) \\
			\cline{2-13}
			{ $\rho$=0.05 }& \multirow{2}{*}{ $10^2$ } & Avg. & \myhighlight{2.00021} & 2.00045 & 1.97849 & 2.00187 & 1.04252 & 1.04313 & \myhighlight{1.0046} & 1.04445 & \myhighlight{0.05464} & 0.26344 \\
			& & (MSE) & (0.06295) & (0.0628) & (0.08027) & (0.05981) & (0.02772) & (0.02748) & (0.02502) & (0.03137) & (0.00154) & (0.04657) \\
			\cline{2-13}
			\hline
			& \multirow{2}{*}{ $10^4$ } & Avg & 1.50027 & 1.50027 & 1.50092 & \myhighlight{1.50023} & 0.99924 & 0.99927 & \myhighlight{0.99952} & 0.99918 & \myhighlight{0.04971} & 0.17777 \\
			{$\beta$=1.5 }& & (MSE) & (0.00037) & (0.00037) & (0.00077) & (0.00037) & (0.00018) & (0.00018) & (0.00022) & (0.00018) & (0.00001) & (0.02054) \\
			\cline{2-13}
			\multirow{2}{*}{ $\gamma$=1 }& \multirow{2}{*}{ $10^3$ } & Avg. & \myhighlight{1.49405} & 1.49405 & 1.48418 & 1.49402 & 1.00938 & 1.00921 & \myhighlight{1.00833} & 1.00914 & \myhighlight{0.05295} & 0.19635 \\
			& & (MSE) & (0.00362) & (0.00362) & (0.00737) & (0.00360) & (0.00180) & (0.00177) & (0.00215) & (0.00180) & (0.00019) & (0.02442) \\
			\cline{2-13}
			{ $\rho$=0.05 }& \multirow{2}{*}{ $10^2$ } & Avg. & 1.41263 & 1.41271 & \myhighlight{1.45518} & 1.43939 & 1.1211 & 1.11986 & \myhighlight{1.07577} & 1.1441 & \myhighlight{0.07833} & 0.27808 \\
			& & (MSE) & (0.0796) & (0.07962) & (0.23004) & (0.04982) & (0.10955) & (0.10715) & (0.07835) & (0.15437) & (0.00533) & (0.05584) \\
			\cline{2-13}
			\hline
		\end{tabular}
\caption{\label{table:bigT}
Summary of the numerical results for the longer cutoff times. Here, the values of $T$ are respectively 9 for $\theta_1$, 6 for $\theta_2$,  and 7 for $\theta_3$ such that the epidemic process almost ended by $T$ (also see \Cref{fig:3SIR}). Method~3 yields accurate estimates without requiring knowledge of the size of the susceptible population~$n$.
}
\end{sidewaystable}

\begin{sidewaystable}[ph!]
		\footnotesize
		\centering
		\begin{tabular}{|c|c|c|c|c|c|c|c|c|c|c|c|c|}
			\hline
			& \multirow{2}{*}{ n } & \multirow{2}{*}{ Statistics } & \multicolumn{4}{|c|}{ $\beta$ } & \multicolumn{4}{|c|}{ $\gamma$ } & \multicolumn{2}{|c|}{$\rho$} \\
			\cline{4-13}
			& & & Method~1 & Method~2 & Method~3 & Method~4 & Method~1 & Method~2 & Method~3 & Method~4 & Method~3 & Method~4\\
			\hline
			& \multirow{2}{*}{ $10^4$ } & Avg. & 2.0443 & 2.0443 & \myhighlight{2.0075} & 2.0442 & 0.4996 & 0.4996 & 0.4995 & \myhighlight{0.4997} & \myhighlight{0.0490} & 0.1960 \\
			{$\beta$=2 }& & (MSE) & (0.00221) & (0.00221) & (0.00093) & (0.00221) & (0.00006) & (0.00006) & (0.00006) & (0.00005) & (0.00001) & (0.02416) \\
			\cline{2-13}
			\multirow{2}{*}{ $\gamma$=0.5 }& \multirow{2}{*}{ $10^3$ } & Avg. & 2.0041 & 2.0040 & 1.9927 & \myhighlight{2.0000} & 0.5034 & 0.5033 & \myhighlight{0.5027} & 0.5035 & \myhighlight{0.0531} & 0.1971 \\
			& & (MSE) & (0.00545) & (5.44670) & (0.00994) & (0.00547) & (0.00067) & (0.00067) & (0.00065) & (0.00068) & (0.00015) & (0.02390) \\
			\cline{2-13}
			{ $\rho$=0.05 }& \multirow{2}{*}{ $10^2$ } & Avg. & 2.0101 & \myhighlight{2.0100} & 2.0313 & 2.0117 & 0.5059 & 0.5069 & \myhighlight{0.5010} & 0.5095 & \myhighlight{0.0574} & 0.2362 \\
			& & (MSE) & (0.07191) & (7.19653) & (0.12782) & (0.06925) & (0.00669) & (0.00677) & (0.00642) & (0.00849) & (0.00184) & (0.03640) \\
			\cline{2-13}

			\hline
			& \multirow{2}{*}{ $10^4$ } & Avg. & 2.1991 & 2.1991 & \myhighlight{2.0131} & 2.1990 & 0.9981 & 0.9982 & \myhighlight{0.9984} & 0.9982 & \myhighlight{0.0489} & 0.2033 \\
			{$\beta$=2 }& & (MSE) & (0.04083) & (0.04083) & (0.00102) & (0.04079) & (0.00031) & (0.00031) & (0.00031) & (0.00031) & (0.00001) & (0.02637) \\
			\cline{2-13}
			\multirow{2}{*}{ $\gamma$=1 }& \multirow{2}{*}{ $10^3$ } & Avg. & \myhighlight{1.9989} & 1.9989 & 1.9959 & 1.9985 & 1.0037 & 1.0036 & \myhighlight{1.0014} & 1.0035 & \myhighlight{0.0508} & 0.1916 \\
			& & (MSE) & (0.00751) & (0.00751) & (0.01117) & (0.00753) & (0.00210) & (0.00210) & (0.00224) & (0.00210) & (0.00012) & (0.02426) \\
			\cline{2-13}
			{ $\rho$=0.05 }& \multirow{2}{*}{ $10^2$ } & Avg. & 1.9979 & \myhighlight{1.9980} & 2.0055 & 2.0040 & 1.0499 & 1.0474 & \myhighlight{1.0241} & 1.0563 & \myhighlight{0.0702} & 0.2433 \\
			& & (MSE) & (0.08047) & (0.08043) & (0.23605) & (0.08257) & (0.11915) & (0.11203) & (0.08004) & (0.14331) & (0.00569) & (0.04042) \\
			\cline{2-13}
			\hline
			& \multirow{2}{*}{ $10^4$ } & Avg. & 1.5713 & 1.5713 & \myhighlight{1.5104} & 1.5715 & 1.0037 & 1.0036 & \myhighlight{1.0032} & 1.00360 & \myhighlight{0.0494} & 0.19920 \\
			{$\beta$=1.5 }& & (MSE) & (0.00804) & (0.00804) & (0.00155) & (0.00809) & (0.00046) & (0.00046) & (0.00046) & (0.00045) & (0.00002) & (0.02496) \\
			\cline{2-13}
			\multirow{2}{*}{ $\gamma$=1 }& \multirow{2}{*}{ $10^3$ } & Avg. & \myhighlight{1.5091} & 1.5091 & 1.5161 & 1.5091 & 1.0079 & 1.0080 & \myhighlight{1.0049} & 1.0085 & \myhighlight{0.0505} & 0.2015 \\
			& & (MSE) & (0.00794) & (0.00794) & (0.01573) & (0.00794) & (0.00381) & (0.00381) & (0.00376) & (0.00389) & (0.00017) & (0.02564) \\
			\cline{2-13}
			{ $\rho$=0.05 }& \multirow{2}{*}{ $10^2$ } & Avg. & 1.4398 & 1.4398 & 1.4762 & \myhighlight{1.4825} & 1.1220 & 1.1192 & \myhighlight{1.0453} & 1.1640 & {\myhighlight{0.0994}} & 0.2380 \\
			& & (MSE) & (0.10451) & (0.10461) & (0.24899) & (0.07014) & (0.16303) & (0.15773) & (0.09018) & (0.23251) & (0.01364) & (0.04004) \\
			\cline{2-13}
			\hline
		\end{tabular}
\caption{\label{table:smallT}%
Summary of the numerical results for the shorter cutoff times. Here, we fix $T=3$ so that the epidemic process is near its peak at $T$ (also see \Cref{fig:3SIR}). Method~3 yields accurate estimates without requiring knowledge of the size of the susceptible population~$n$.
}
\end{sidewaystable}

For the longer cutoff times, \Cref{table:bigT} provides a  summary of the simulation study for the three parameter sets and different initial number of susceptibles $n$. Here, the values of $T$ are respectively 9 for $\theta_1$, 6 for $\theta_2$,  and 7 for $\theta_3$,  that is,  in each case  the epidemic  is almost at its end  by time $T$ (see \Cref{fig:3SIR}). The first four columns show the estimates of $\beta$ according to the Methods 1, 2, 3 and 4. Similarly, the next four columns show estimates  of $\gamma$. The last two columns are reserved for Method 3 and Method 4 estimates of $\rho$. Recall that  $\rho$ is known exactly for Method 1 and Method 2.  
The rows of the table are divided into three parts corresponding to the three settings of the parameter values $\theta_1, \theta_2$, and $\theta_3$. Each of three parts is further segregated into three different classes corresponding to the three different susceptible population sizes $n=10^2, 10^3$, and $10^3$. Finally, in each cell, we show the average of 100 posterior means and \ac{MSE} of parameter estimators. As we can see, the Method~3 based on the \ac{SDS}-likelihood yields accurate estimates for all three parameters $\beta, \gamma$, and $\rho$ even for relatively small values of $n$ (see the results for  $n=10^2$). As one would expect, the \acp{MSE} decrease with increase in $n$ across the four different methods. In particular, our Method~3 has a slightly higher variance than the other methods. However, it is important to note that Method~3 does \emph{not} require knowledge of $n$ whereas the other methods do. Most notable is Method~3's ability to estimate $\rho$ accurately, specially when pitted against the poor performance of Method~4 based on the Gaussian likelihood.


The only difference between  results in  \Cref{table:bigT,table:smallT} are in cutoff times.  Whereas in \Cref{table:bigT} we consider data collected for  most of the epidemic duration (cutoff $T$ is close to the end of the epidemic),  in  \Cref{table:smallT} we consider data with short cutoff $T=3$ that is close  to epidemic peak times.   See \Cref{fig:3SIR} in \Cref{appendix:inference} for a visualization of the \ac{SIR} curves corresponding to these three parameter settings truncated at $T=3$ by a vertical line. Since the inference is based on the heavily truncated data, the
\acp{MSE} in \Cref{table:smallT} are expectedly worse than those in \Cref{table:bigT}. Also, the sharp decrease in \acp{MSE} with increasing $n$ in \Cref{table:bigT} is less pronounced in \Cref{table:smallT}. Nevertheless, the estimates obtained are still quite accurate. Also, the \acp{MSE} for Method~3 are slightly better than those of Method 1 or 2. Interestingly, the parameter $\rho$ is almost always better estimated by Method~3.

Further supplementary numerical results and explanations are provided in  \Cref{appendix:inference}.

\setcounter{equation}{0}
\section{Discussion}
\label{sec:conclusion}
In this paper, we presented a new way of looking at the classical \ac{SIR}-type epidemic models. Our method addresses all the four problems of the classical \ac{SIR} model  identified in \Cref{sec:intro}. Parameter estimation based on the \ac{SDS}-likelihood (described in \Cref{sec:inference}) does not require the effective population size~$n$, addressing problem~\ref{problem1}. The \ac{SDS}-likelihood approach, being a direct consequence of the \ac{SDS} interpretation of the \ac{SIR} \Cref{eq:SIR_ODE}, provides a principled way of specifying the likelihood from epidemiological field data where the effective population size is unknown but large, addressing problem~\ref{problem4}. Although in the current work we do not explicitly illustrate  this, it should be clear that the independence of the individuals' contributions to   \ac{SDS}-likelihood addresses also  the problem of aggregation over individuals (problem~\ref{problem2}) and over time (problem~\ref{problem3}).

It is worth mentioning that, under the \ac{SDS}-likelihood approach, it typically suffices to have much smaller sample of transition data than other inference methods, such as the exact likelihood or the Gaussian likelihood methods. Due to the asymptotic independence of infection and recovery times of individuals (see \Cref{sec:mtm}), the \ac{SDS}-likelihood takes a particularly simple form facilitating a convenient  implementation of a suitable \ac{MCMC} scheme. For ready usage of our method, we have made our code implementation publicly available~\cite{software}.

The proposed method can be readily extended to accommodate   a wide class of  \acp{MTM}. The classical \ac{SIR} model has been chosen here merely as an  example to illustrate the ideas underpinning the \ac{SDS} interpretation of unidirectional \acp{MTM}. Indeed, the machinery developed in the present paper goes beyond \ac{SIR} models, and it can be immediately applied to  more general epidemic processes as well as general \acp{MTM} arising in physics and chemistry.  In particular, we believe the \ac{SDS} tools can be applied to certain subclasses of \acp{CRN} models in which the individual  species molecules can be tracked as they undergo chemical reactions.

In many studies of epidemiological field data, the effective population size is assumed to be very large. For instance, a total population size of $10^6$ was assumed in \cite{althaus2014estimating,getz2018discrete}. Our method is particularly appropriate for such settings. However, since our method hinges on a \acl{LLN} (see \Cref{sec:background}), the rate of convergence of the scaled processes to the \ac{LLN} limit, which coincides with the \ac{SIR} \Cref{eq:SIR_ODE}, is crucial for the quality of inference based on the \ac{MTM}-likelihood. Therefore, we need to establish a \acl{LDP} for the scaled processes.  This is particularly important for small-scale epidemics. Even though our numerical results are encouraging for values of $n$ as small as 100, quantifying the rate of convergence will be useful. We did not consider an \ac{LDP} in this paper, but we believe standard techniques \cite{dembo2010LargeDeviations,feng2006large,pardoux2017large,dolgoarshinnykh2009sample,Dupius2011LDP,pardoux2016ldp_Poisson} can be used for this purpose. Another direction of future investigation will be to consider non-mass-action systems and, eventually,  non-Markovian systems with non-exponential holding times. Note that the original Sellke construction does not assume Markovianness of the stochastic system. The Markovian version presented in \Cref{sec:background} has been adapted to our context.

For many epidemiological scenarios, the mass-action assumption is untenable. Several network-based models have been proposed in the recent times \cite{Volz2008,kenah2007secondLook,Newman2002network}.   Asymptotic study of those models in the form of various large-graph limits has also been done \cite{Burch2017network,jacobsen2018LLN,Khudabukhsh2017FCLT}.  Therefore, extending our method to network-based models appears to be a  natural next step  that we hope  to take in the near future.




\begin{appendices}
\renewcommand{\theequation}{\thesection.\arabic{equation}}

  \crefalias{section}{appsec}
  \setcounter{equation}{0}
  \section{Mathematical background}
\label{appendix:prelims}
\subsection{Lumpability of a Markov chain}
The (strong) lumpability\footnote{There is also a notion of weak lumpability in the theory of Markov processes.} of a \ac{CTMC}   can be  described in terms of lumpability of a linear system of \acp{ODE}. Consider the linear system $  \dot{y} =  y A  $,
where $A = ((a_{i,j}))$ is a $K\times K$ matrix (representing the transition rate or the infinitesimal generator matrix  of the corresponding \ac{CTMC} on state space $\mathcal{Y}\defeq \{1,2,\ldots,K\}$). 
\begin{myDefinition}[Lumpability of a linear system \cite{khudabukhsh2018Lumpability,Simon2011Exact}]
  The linear system $  \dot{y} = y A  $ is said to be lumpable with respect to a partition $\{  \mathcal{Y}_1, \mathcal{Y}_2,\ldots, \mathcal{Y}_M \}$ of $\mathcal{Y}$, if there exists an $M\times K$ matrix $B = ((b_{i,j}))$ satisfying Dynkin's criterion (\ie, if $b_{i,j} = \sum_{l \in \mathcal{Y}_j } a_{u,l}  =  \sum_{l  \in \mathcal{Y}_j } a_{v,l} $ for all $u, v \in \mathcal{Y}_i$).  The matrix $B$ is often called a lumping of $A$.  The following is immediate: If $B$ is a lumping of $A$, then there exists an $K\times M$ matrix $V$ such that $AV=V B$. \QEDB
\end{myDefinition}

\subsection{Gaussian likelihood}
\label{appendix:gaussian}
For large $n$  and  $m=\rho n$ we can replace the exact likelihood \Cref{eq:lkh1}  with the approximate Gaussian one  because the score processes $\partial \ell_1(\theta\mid \cdot)/\partial \beta$ and  $\partial \ell_1(\theta\mid \cdot)/\partial \ga$ are asymptotically ($n\to \infty$) independent and Gaussian.  This gives the log-likelihood formula   
\begin{align}\label{eq:lkh3} 
\ell_3(\theta\mid X(t)_{t\in (0,T]})= & -  (n(1-s_T))^{-1} \left[z_I-\beta\int_0^T S(t)I(t)\, \differential{t}/n\right]^2   \nonumber  \\
&- (n (1+\rho-s_T-\iota_T))^{-1} \left[z_R-\gamma \int_0^T I(t)\, \differential{t}\right]^2 \nonumber\\ &- \log\left(2\pi(1-s_T)(1+\rho-s_T-\iota_T) \right) \eqcomma
\end{align}
 where $(s_t,\iota_t)$ are the trajectories of the \ac{ODE} system \Cref{eq:mean_field_SIR}. Note that maximization of $\ell_3(\theta \mid \cdot)$ also leads to  \acp{MLE}  \Cref{eq:mle1}. However, maximization with respect to the parameter $\rho$ needs to be done implicitly by adjusting the deterministic trajectory  $(s_t,\iota_t)$.  Essentially, this boils down to maximizing the third term in \Cref{eq:lkh3} because the \acp{MLE} of $\beta$ and $\gamma$ are found by setting the squared terms (the first two terms in  \Cref{eq:lkh3}) to zero. A drawback of the Gaussian likelihood $\ell_3$  is that the accuracy of the estimate for $\rho$ may be poor, particularly when $n$ is not large,  because the third term depends only on the solution of the \ac{ODE} system \Cref{eq:mean_field_SIR} and not on sample data.  

   \setcounter{equation}{0}
   \section{Additional numerical results}
\label{appendix:inference}
Here, we provide additional numerical results. In particular, we show the posterior plots and crucial diagnostic statistics for the \ac{MCMC} methods.

\begin{figure}[htp]
	\begin{subfigure}[h]{0.32\textwidth}
		\includegraphics[width=\textwidth]{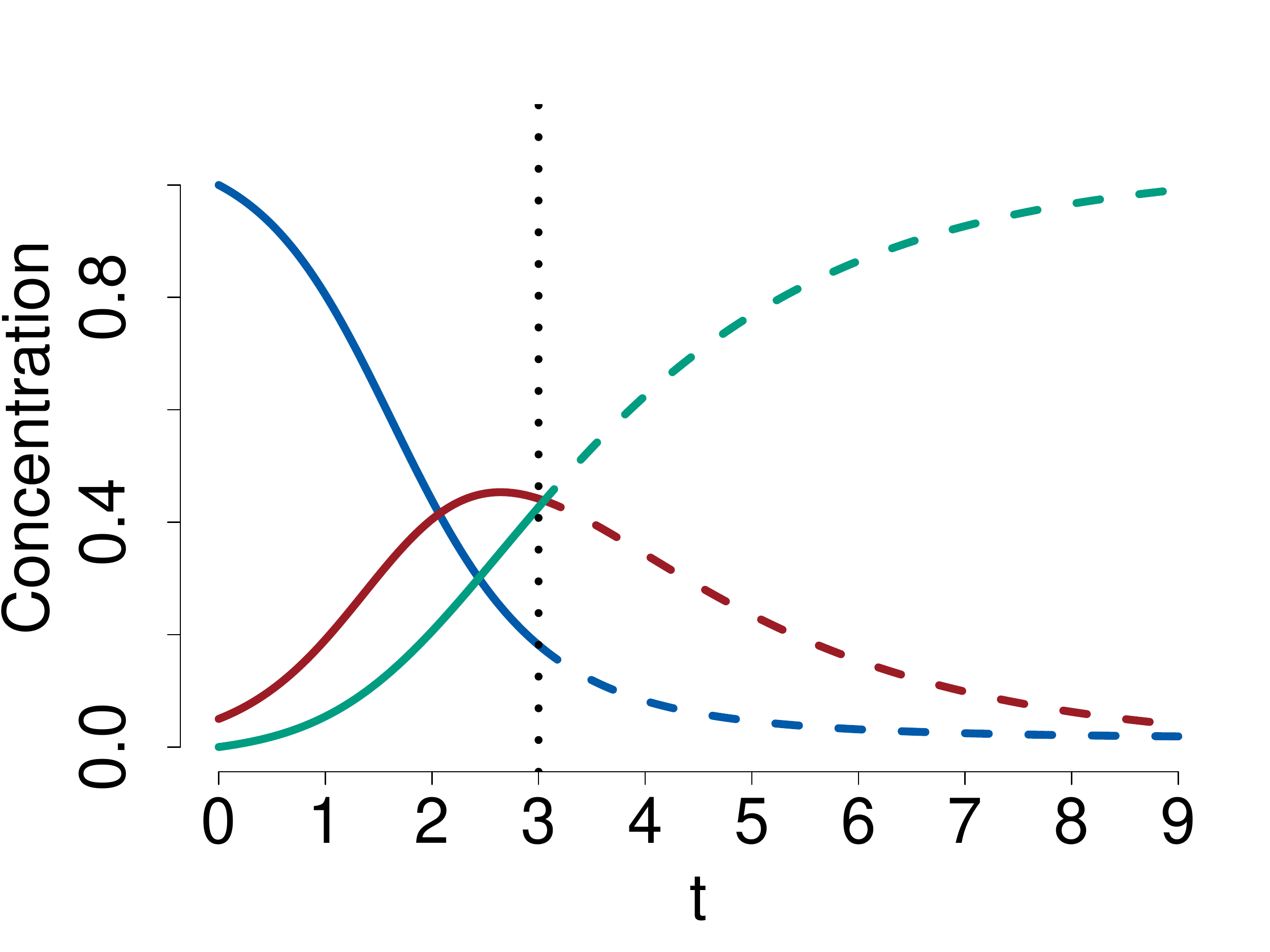}
		\caption{$\beta$=2, $\gamma$=0.5, $\rho$=0.05}
	\end{subfigure}
	\begin{subfigure}[h]{0.32\textwidth}
		\includegraphics[width=\textwidth]{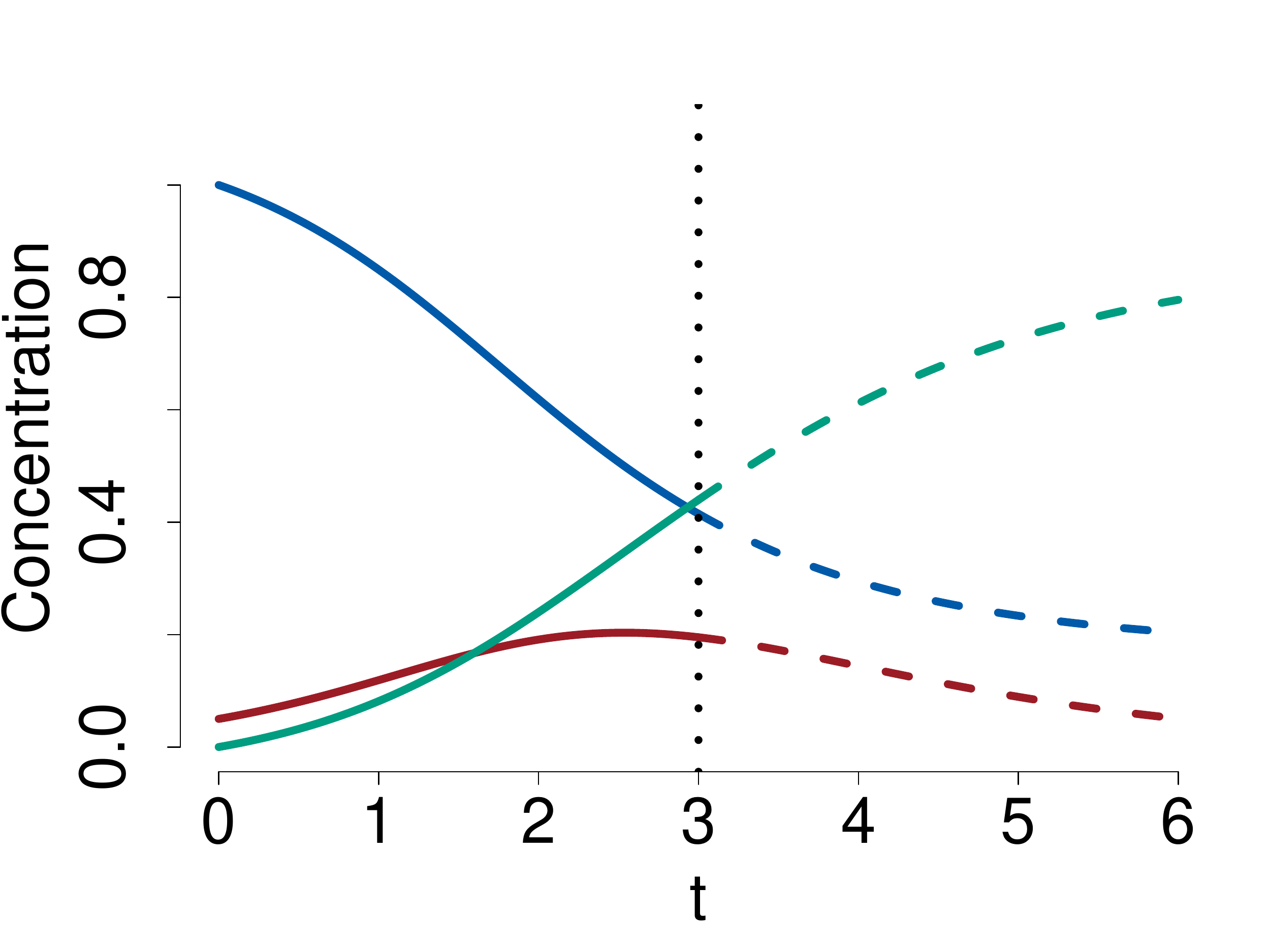}
		\caption{$\beta$=2, $\gamma$=1.0, $\rho$=0.05}
	\end{subfigure}
	\begin{subfigure}[h]{0.32\textwidth}
		\includegraphics[width=\textwidth]{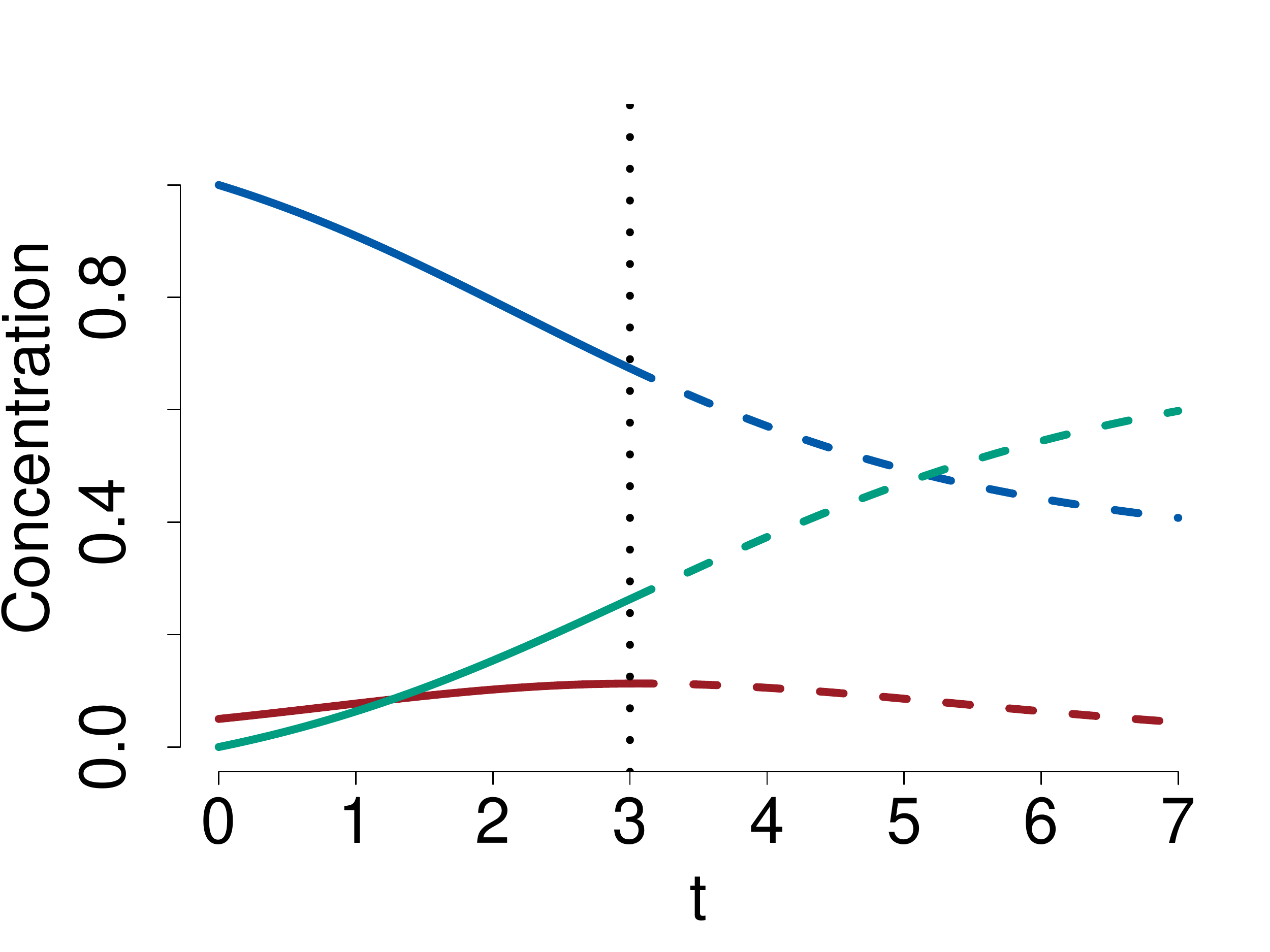}
		\caption{$\beta$=1.5, $\gamma$=1.0, $\rho$=0.05}
	\end{subfigure}
	\caption{\label{fig:3SIR}%
The \ac{SIR}  curves for the three different parameter values considered in \Cref{sec:numerical}. The initial values are $S_0 = 1$, $R_0 = \rho$, and $R_0 = 0$. The vertical dotted lines represent the cutoff time.
}
\end{figure}

The cutoff times are decided based on \Cref{fig:3SIR}. The idea is to study the impact of censoring on the quality of the inference procedure. Therefore, for each parameter setting, we choose two cutoff times: one near the peak of the epidemic and one near the end of the epidemic. The vertical lines in  \Cref{fig:3SIR} demarcates the smaller cutoff times for each of the three settings of the parameter values.

\begin{figure}[htp]
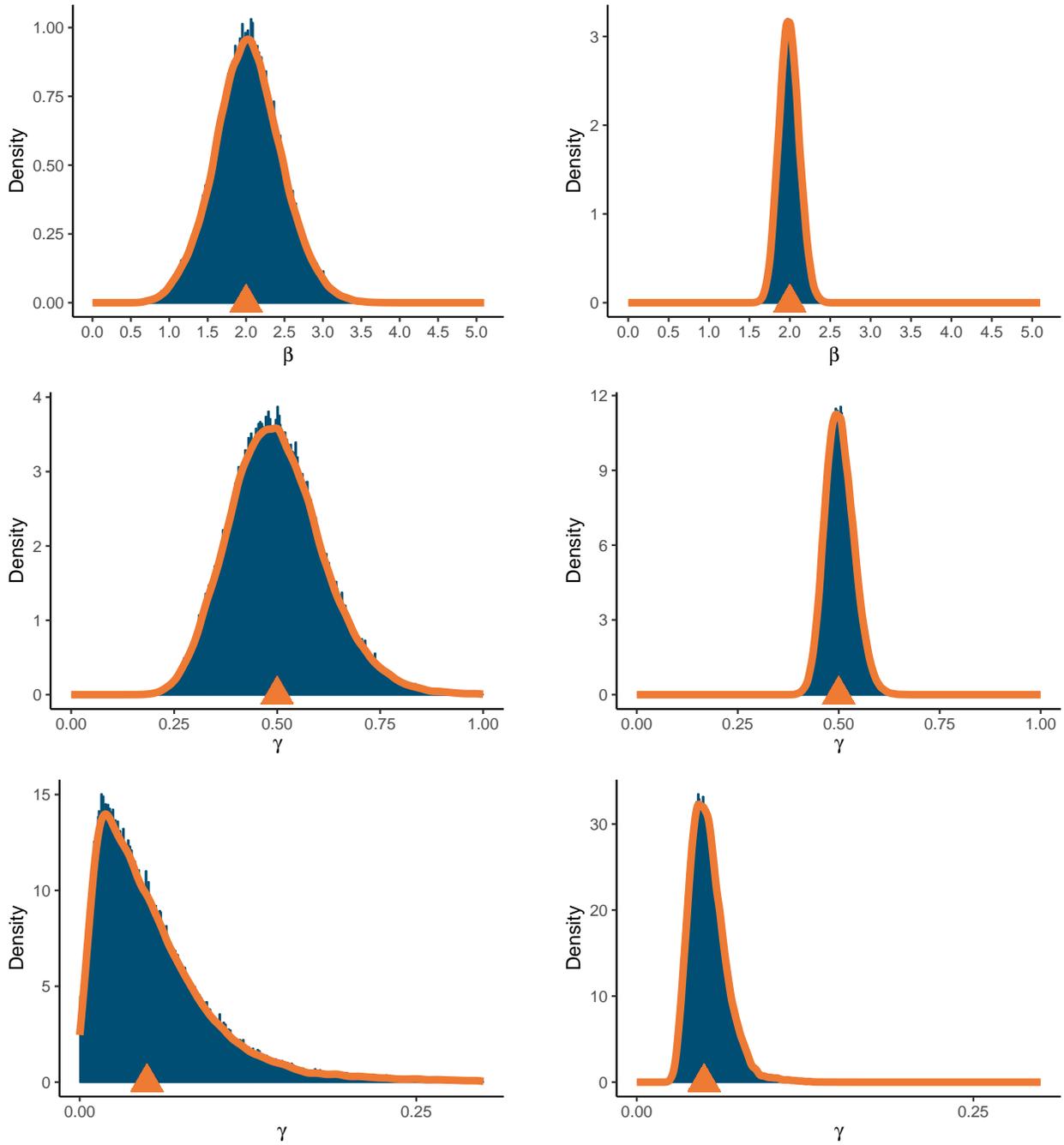

\centering
\begin{subfigure}{0.48\columnwidth}
\includegraphics[width=\columnwidth]{{{beta_posterior_100_2_0.5_0.05_3}}}
\end{subfigure}\hfill
\begin{subfigure}{0.48\columnwidth}
\includegraphics[width=\columnwidth]{{{beta_posterior_1000_2_0.5_0.05_3}}}
\end{subfigure}
\begin{subfigure}{0.48\columnwidth}
\includegraphics[width=\columnwidth]{{{gamma_posterior_100_2_0.5_0.05_3}}}
\end{subfigure}\hfill
\begin{subfigure}{0.48\columnwidth}
\includegraphics[width=\columnwidth]{{{gamma_posterior_1000_2_0.5_0.05_3}}}
\end{subfigure}
\begin{subfigure}{0.48\columnwidth}
\includegraphics[width=\columnwidth]{{{rho_posterior_100_2_0.5_0.05_3}}}
\end{subfigure}
\hfill
\begin{subfigure}{0.48\columnwidth}
\includegraphics[width=\columnwidth]{{{rho_posterior_1000_2_0.5_0.05_3}}}
\end{subfigure}
\caption{\label{fig:posterior_setting1_smallT}%
The posterior distributions of the Method~3 estimators of $\beta, \gamma$, and $\rho$ based on the \ac{SDS}-likelihood for the smaller cutoff time case ($T=3$).  The panels in the first row correspond to $n=10^2$, and those in the second row (bottom) correspond to $n=10^3$.  The true parameter values are $\beta=2,\gamma=0.5,\rho=0.05$ (parameter setting~1).
}
\end{figure}

\begin{figure}[htp]
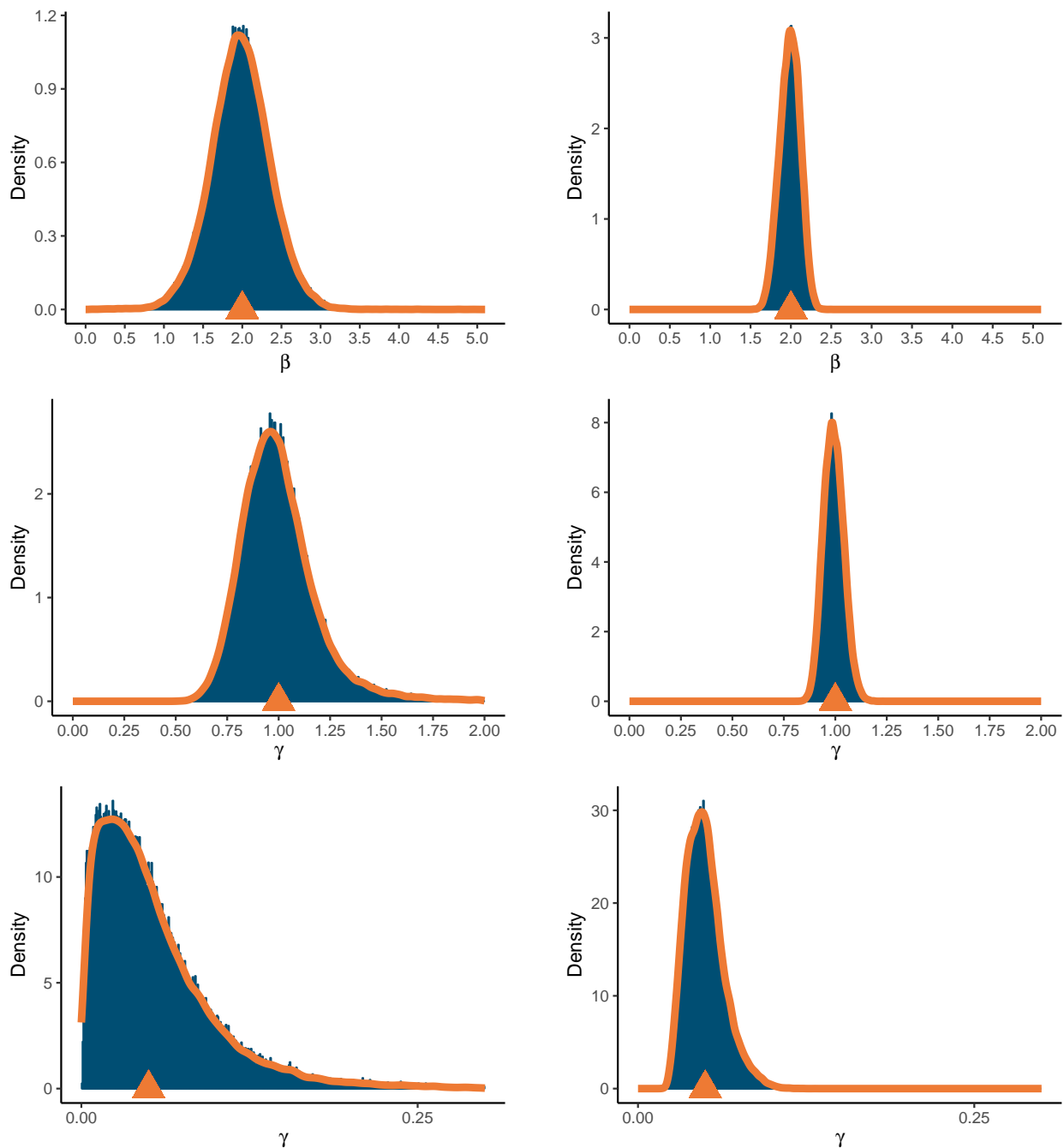

\centering
\begin{subfigure}{0.48\columnwidth}
\includegraphics[width=\columnwidth]{{{beta_posterior_100_2_1_0.05_6}}}
\end{subfigure}\hfill
\begin{subfigure}{0.48\columnwidth}
\includegraphics[width=\columnwidth]{{{beta_posterior_1000_2_1_0.05_6}}}
\end{subfigure}
\begin{subfigure}{0.48\columnwidth}
\includegraphics[width=\columnwidth]{{{gamma_posterior_100_2_1_0.05_6}}}
\end{subfigure}\hfill
\begin{subfigure}{0.48\columnwidth}
\includegraphics[width=\columnwidth]{{{gamma_posterior_1000_2_1_0.05_6}}}
\end{subfigure}
\begin{subfigure}{0.48\columnwidth}
\includegraphics[width=\columnwidth]{{{rho_posterior_100_2_1_0.05_6}}}
\end{subfigure}
\hfill
\begin{subfigure}{0.48\columnwidth}
\includegraphics[width=\columnwidth]{{{rho_posterior_1000_2_1_0.05_6}}}
\end{subfigure}
\caption{\label{fig:posterior_setting2_largeT}%
The posterior distributions of the Method~3 estimators of $\beta, \gamma$, and $\rho$ based on the \ac{SDS}-likelihood for the larger cutoff time case ($T=6$).  The panels in the first (left) column  correspond to $n=10^2$, and those in the second (right)  column correspond to $n=10^3$.  The true parameter values are $\beta=2,\gamma=1,\rho=0.05$ (parameter setting~2).
}
\end{figure}

\begin{figure}[htp]
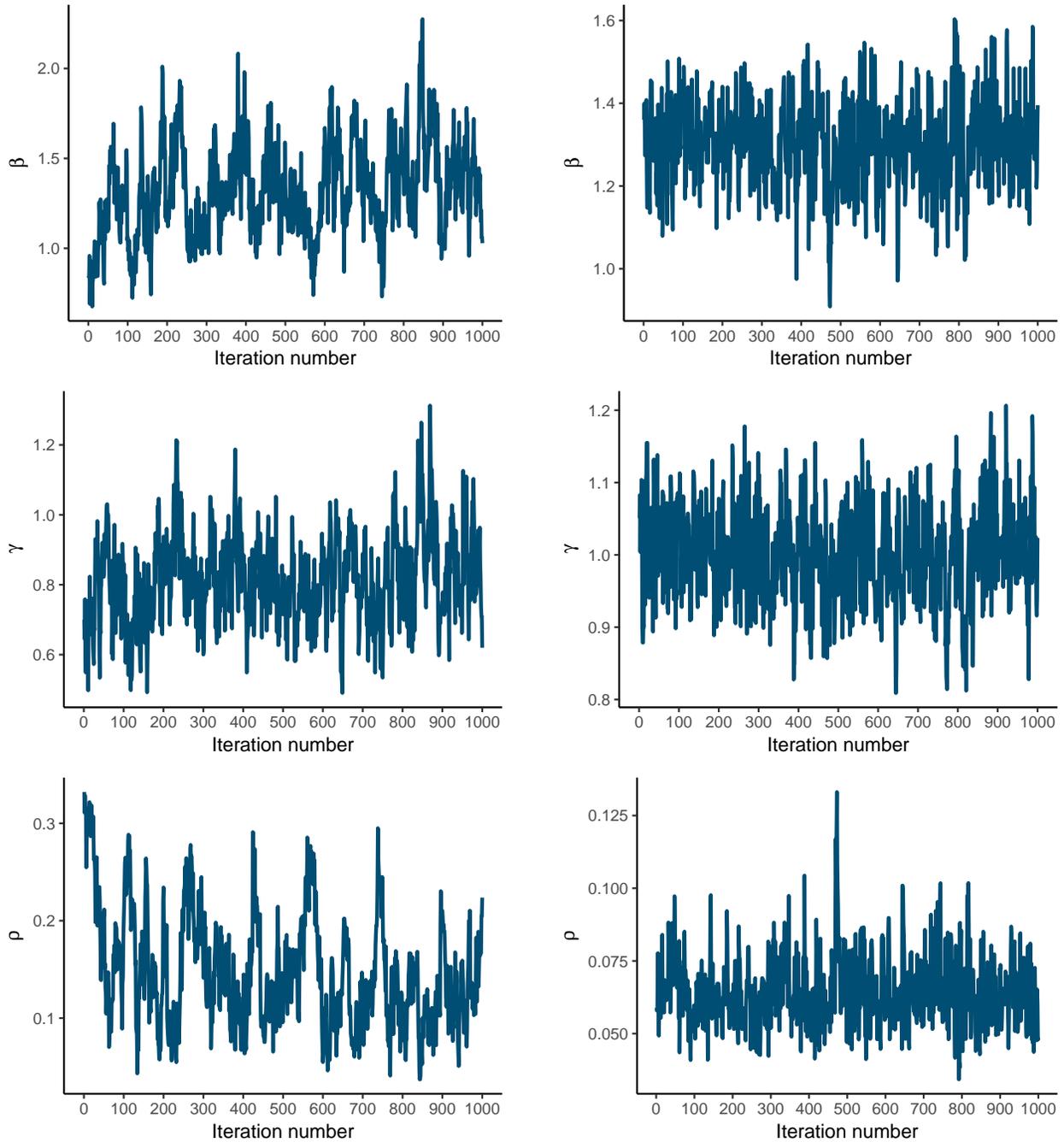

\centering
\begin{subfigure}{0.48\columnwidth}
\includegraphics[width=\columnwidth]{{{beta_chain_100_1.5_1_0.05_3}}}
\end{subfigure}\hfill
\begin{subfigure}{0.48\columnwidth}
\includegraphics[width=\columnwidth]{{{beta_chain_1000_1.5_1_0.05_3}}}
\end{subfigure}
\begin{subfigure}{0.48\columnwidth}
\includegraphics[width=\columnwidth]{{{gamma_chain_100_1.5_1_0.05_3}}}
\end{subfigure}\hfill
\begin{subfigure}{0.48\columnwidth}
\includegraphics[width=\columnwidth]{{{gamma_chain_1000_1.5_1_0.05_3}}}
\end{subfigure}
\begin{subfigure}{0.48\columnwidth}
\includegraphics[width=\columnwidth]{{{rho_chain_100_1.5_1_0.05_3}}}
\end{subfigure}
\hfill
\begin{subfigure}{0.48\columnwidth}
\includegraphics[width=\columnwidth]{{{rho_chain_1000_1.5_1_0.05_3}}}
\end{subfigure}
\caption{\label{fig:chain_setting3_smallT}%
  The (thinned) trace of a single Markov chain in the \ac{MCMC} implementation of Method~3 for the smaller cutoff time ($T = 3$). Separate panels are shown for each of the parameters $\beta, \gamma$, and $\rho$. The panels in the left column  correspond to $n=10^2$, and those in the right column correspond to $n=10^3$.  The true parameter values are $\beta=1.5$, $\gamma=1$, and $\rho=0.05$ (parameter setting~3).
}
\end{figure}

In \Cref{fig:posterior_setting1_smallT,fig:posterior_setting2_largeT},  we show the posterior distributions of the Method~3 estimators of $\beta, \gamma$, and $\beta$ based on the \ac{SDS}-likelihood. To avoid repetition, we show only two posterior plots: one for the parameter setting $\theta_1$ for the smaller cutoff time case in \Cref{fig:posterior_setting1_smallT} and one for the parameter setting $\theta_2$ for the larger cutoff time case in \Cref{fig:posterior_setting2_largeT}. As shown in \Cref{table:smallT,table:bigT}, the variance of the posterior distributions shrink drastically as we increase $n$ from $10^2$ to $10^3$. We do not show the posterior distributions for the $n=10^4$ case because it does not provide any additional insights into the quality of the inference procedure except for the fact that the posterior variance further reduces. Finally, we provide additional diagnostic statistics for the \ac{MCMC} implementation of Method~3  in \Cref{fig:chain_setting3_smallT}. We show the (thinned) trace of a single Markov chain for two different values of $n$, namely $n=10^2$, and $10^3$. As \Cref{fig:chain_setting3_smallT} shows, the chain mixes faster when $n=10^3$ than when $n=10^2$. This is expected because Method~3 is essentially based on an \ac{LLN} of the scaled Poisson processes keeping track of the population counts. As before, we omit the trace plots corresponding to the $n=10^4$ case. For completeness, we consider the third parameter setting $\theta_3$ in \Cref{fig:chain_setting3_smallT}. To avoid repetition, we do not show trace plots for the other parameter settings. Nevertheless, the Markov chains converge for the other parameter settings as well.


   \section{Acronyms}\label{appendix:acronym}

\begin{acronym}[OWL-QN]
	\acro{ABM}{Agent-based Model}
	\acro{ADMM}{Alternating Direction Method of Multipliers}
	\acro{BA}{Barab\'asi-Albert}
	\acro{BCS}{Bioinspired Communication Systems}
	\acro{BM}{Brownian Motion}
	\acro{CBQA}{Cost-Based Queue-Aware}
	\acro{CBS}{Cost-Based Scheduling}
	\acro{CCDF}{Complementary Cumulative Distribution Function}
	\acro{CDC}{Centers for Disease Control and Prevention}
	\acro{CDF}{Cumulative Distribution Function}
	\acro{CDN}{Content Distribution Network}
	\acro{CIM}{Conditional Intensity Matrix}
	\acro{CLT}{Central Limit Theorem}
	\acro{CM}{Configuration Model}
	\acro{CME}{Chemical Master Equation}
	\acro{CoM}{Compartmental Model}
	\acro{CRC}{Collaborative Research Centre}
	\acro{CRM}{Conditional Random Measure}
	\acro{CRN}{Chemical Reaction Network}
	\acro{CTBN}{Continuous Time Bayesian Network}
	\acro{CTMC}{Continuous Time Markov Chain}
	\acro{DCFTP}{Dominated Coupling From The Past }
	\acro{DFG}{German Research Foundation}
	\acro{DTMC}{Discrete Time Markov Chain}
\acro{DRC}{Democratic Republic of Congo}
	\acro{ECMP}{Equal-cost Multi-path routing}
	\acro{EDF}{Earliest Deadline First}
	\acro{ER}{Erd\"{o}s-R\'{e}nyi}
	\acro{ESI}{Enzyme-Substrate-Inhibitor}
	\acro{FCFS}{First Come First Served}
	\acro{FCLT}{Functional Central Limit Theorem}
	\acro{FIFO}{First In First Out}
	\acro{FJ}{Fork-Join}
	\acro{GBP}{General Branching Process}
	\acro{ID}{Information-Dissemination}
	\acro{iid}{independent and identically distributed}
	\acro{IoT}{Internet of Things}
	\acro{IPS}{Interacting Particle System}
	\acro{IT}{Information Technology}
	\acro{JIQ}{Join-Idle-Queue}
	\acro{JMC}{Join the Minimum Cost}
	\acro{JSQ}{Join the Shortest Queue}
	\acro{KL}{Kullback-Leibler}
	\acro{LDF}{Latest Deadline First}
	\acro{LDP}{Large Deviations Principle}
	\acro{LLN}{Law of Large Numbers}
	\acrodefplural{LLN}[LLNs]{Laws of Large Numbers}
	\acro{LNA}{Linear Noise Approximation}
	\acro{MABM}{Markovian Agent-based Model}
	\acro{MAKI}{Multi-Mechanism Adaptation for the Future Internet}
	\acro{MAPK}{Mitogen-activated Protein Kinase}
	\acro{MCMC}{Markov Chain Monte Carlo}
	\acro{MDS}{Maximum Distance Separable}
	\acro{MGF}{Moment Generating Function}
	\acro{MLE}{Maximum Likelihood Estimate}
	\acro{MM}{Michaelis-Menten}
	\acro{MPI}{Message Passing Interface}
	\acro{MPTCP}[Multi-path TCP]{Multi-path Transmission Control Protocol}
	\acro{MTM}{Mass Transfer Model}
	\acro{MSE}{Mean Squared Error}
	\acro{ODE}{Ordinary Differential Equation}
	\acro{P2P}{Peer-to-Peer}
	\acro{PDE}{Partial Differential Equation}
	\acro{PDF}{Probability Density Function}
	\acro{PGF}{Probability Generating Function}
	\acro{PGM}{Probabilistic Graphical Model}
	\acro{PMF}{Probability Mass Function}
	\acro{psd}{positive semi-definite}
	\acro{PT}{Poisson-type}
	\acro{QoE}{Quality of Experience}
	\acro{QoS}{Quality of Service}
	\acro{QSSA}{Quasi-Steady State Approximation}
\acro{RBM}{Reflecting Brownian Motion}
	\acro{rQSSA}{reversible QSSA}
	\acro{SAN}{Stochastic Automata Network}
	\acro{SD}{Standard Deviation}
	\acro{SDS}{Survival Dynamical System}
	\acro{SEIR}{Susceptible-Exposed-Infected-Recovered}
	\acro{SI}{Susceptible-Infected}
	\acro{SIR}{Susceptible-Infected-Recovered}
	\acro{SIS}{Susceptible-Infected-Susceptible}
	\acro{sQSSA}{standard QSSA}
	\acro{SRBM}{Semi-martingale Reflecting Brownian Motion}
	\acro{SRPT}{Shortest Remaining Processing Time}
	\acro{ssLNA}{Slow-scale Linear Noise Approximation}
	\acro{STC}{Stone Throwing Construction}
	\acro{TCP}{Transmission Control Protocol}
	\acro{tQSSA}{total QSSA}
	\acro{WS}{Watts-Strogatz}
	\acro{whp}{with high probability}
	\acro{WSU}{Washington State University}
	\acro{RAM}{Robust Adaptive Metropolis}
	\acro{ASM}{Adaptive Scaling Metropolis}
\end{acronym}

\section*{Competing interests}
The authors declare  no competing interests.

\section*{Authors' contributions}
GAR and EK conceived and designed the research. BC provided  numerical examples, contributed analysis tools, and helped write the paper.  WKB and GAR wrote the paper. All authors helped in editing and proofreading the final manuscript.

\section*{Acknowledgments}
 The large part of this research was conducted during  the Mathematical Biosciences Institute (MBI) semester-long program on modeling infectious diseases in Spring 2018. The authors would like to thank MBI and its staff for their hospitality.

\section*{Funding}
BC was supported by the National Research Foundation of Korea (NRF) grant NRF-2017R1D1A3B03031008. GAR was supported by the National Science Foundation (NSF) under grants NSF-DMS 1440386 and NSF-DMS 1513489. EK was supported by the National Institute of General Medical Sciences (NIGMS) grant U54 GM111274. EK and WKB were supported by the National Institute of Allergy and Infectious Diseases (NIAID) grant R01 AI116770. The content is solely the responsibility of the authors and does not represent the official views of NRF, NSF, NIGMS, or NIAID.
\end{appendices}

\bibliographystyle{BibStyle_WKB}

\bibliography{bibliography_WKB,bibliography_GR,bibliography_EK,bibliography_BC,add_bibl}

\end{document}